\documentclass[journal,10pt]{IEEEtran}

\usepackage[cmex10]{amsmath}
\usepackage{mathrsfs}
\usepackage{mathtools}

\usepackage{colortbl}
\usepackage[dvipsnames]{xcolor}

\usepackage{amsmath}
\usepackage{amssymb}
\usepackage{stfloats}
\usepackage{cite}
\usepackage{graphicx}
\usepackage{subfigure}
\usepackage{bm}
\usepackage{pifont}
\usepackage{algorithmic}
\usepackage[procnumbered,ruled]{algorithm2e}
\usepackage[normalem]{ulem}
\newtheorem{theorem}{\it Theorem}

\newtheorem{lemma}{\it Lemma}

\begin{document}

\title{Performance Analysis of Spatiotemporal 2-D Polar Codes for Massive MIMO with MMSE Receivers}

\author{Yaqi~Li,~\IEEEmembership{Student Member,~IEEE,}
	    ~Xiaohu~You,~\IEEEmembership{Fellow,~IEEE,}
        ~Jiamin~Li,~\IEEEmembership{Member,~IEEE,}
        ~Chen~Ji,
        ~and~Bin~Sheng,~\IEEEmembership{Member,~IEEE}

\thanks{This work was supported in part by Major Science and Technology Project of Jiangsu Province under Grant BG2024002, by the National Natural Science Foundation of China (NSFC) under Grants 62331009, by the Fundamental Research Funds for the Central Universities under Grant 2242022k60006, and by the Major Key Project of PCL (PCL2021A01-2). \emph{(Corresponding Author: Xiaohu You)}}

\thanks{Y. Li, X. You, J. Li and B. Sheng are with the National Mobile Communications Research Laboratory, Southeast University, Nanjing 210096, China (e-mail:{230258100, xhyu, jiaminli, sbdtt}@seu.edu.cn). X. You, J. Li and B. Sheng are also with Purple Mountain Laboratories,
Nanjing, 211111, China.}

\thanks{C. Ji is with School of Information Science and Technology, Nantong University, Nantong, 226019, China. (email: gwidjin@ntu.edu.cn)}

}

\maketitle

\begin{abstract}

	With the evolution from 5G to 6G, ultra-reliable low-latency communication (URLLC) faces increasingly stringent performance requirements. Lower latency constraints demand shorter channel coding lengths, which can severely degrade decoding performance. The massive multiple-input multiple-output (MIMO) system is considered a crucial technology to address this challenge due to its abundant spatial degrees of freedom (DoF). While polar codes are theoretically capacity-achieving in the limit of infinite code length, their practical applicability is limited by significant decoding latency. In this paper, we establish a unified theoretical framework and propose a novel spatiotemporal two-dimensional (2-D) polar coding scheme for massive MIMO systems employing minimum mean square error (MMSE) receivers. The polar transform is jointly applied over both spatial and temporal dimensions to fully exploit the large spatial DoF. By leveraging the near-deterministic signal-to-interference-plus-noise ratio (SINR) property of MMSE detection, the spatial domain is modeled as a set of parallel Gaussian sub-channels. Within this framework, we perform a theoretical analysis of the 2-D polarization behavior using the Gaussian approximation method, and the capacity-achieving property of the proposed scheme is proved under finite blocklength constraints and large spatial DoF. Simulation results further demonstrate that, compared to traditional time-domain polar codes, the proposed 2-D scheme can significantly reduce latency while guaranteeing reliability, or alternatively improve reliability under the same latency constraint---offering a capacity-achieving and latency-efficient channel coding solution for massive MIMO systems in future 6G URLLC scenarios.
	
\end{abstract}

\begin{IEEEkeywords}
	Polar coding, spatiotemporal, Gaussian approximation, capacity-achieving, Massive MIMO
\end{IEEEkeywords}

\IEEEpeerreviewmaketitle

\section{Introduction}\label{sec1}
\IEEEPARstart{A}{s} one of the core communication scenarios of the fifth generation (5G) mobile communication, Ultra-reliable low-latency communication (URLLC) provides technical support for mission-critical applications such as telemedicine, virtual reality, and autonomous driving, which have high requirements for end-to-end delay and reliability \cite{schulz2017latency,sutton2019enabling,sachs20185g}. However, with the increasingly rigorous requirements of these new applications, how to achieve high data rate transmission while ensuring high reliability and low latency has emerged as a critical challenge demanding urgent solutions. In 5G communications, short blocklength can be used to reduce latency, but it leads to a significant degradation in decoding performance, compromising transmission reliability. For the upcoming sixth generation (6G), the performance requirements on latency and reliability are further tightened to meet the 6G TK$\mu$ extreme connectivity \cite{you2023toward,saad2019vision}. Therefore, the role of URLLC will be further enhanced in the 6G era. This puts forward higher demands for channel coding design, and new coding schemes that can flexibly balance coding performance and code length are highly required.

The classical Shannon information theory reveals the performance tradeoff between information rate, latency, and reliability for asymptotically long codewords. Nevertheless, it does not fully characterize the latency-reliability trade-off for short blocklength coding \cite{1998Constrained,2007A,2010Channel}. In multiple-input multiple-output (MIMO) systems with finite blocklength, the achievable rate deviates rapidly from the theoretical limit of channel capacity as the blocklength decreases, resulting in the channel capacity collapse effect \cite{you20236g}. To address this problem, \cite{you2023closed} illustrates that the performance of MIMO systems can be improved under fixed blocklength by increasing the spatial degrees of freedom (DoF) of MIMO systems. Therefore, massive MIMO systems equipped with a large number of antennas are expected to overcome the performance limitations imposed by the traditional latency-reliability tradeoff to realize URLLC. Furthermore, based on the spectral analysis of the minimum mean square error (MMSE) receiver in massive MIMO systems, the signal-to-interference-plus-noise ratio (SINR) converges to a normal distribution with decreasing variance, under the asymptotic condition that the scale of antennas tends to infinity \cite{hoydis2011iterative,li2005distribution,kammoun2009central,moustakas2013sinr}. This deterministic approximation property indicates that the massive MIMO channel can be approximately equivalent to a large set of parallel Gaussian channels with consistent SINR. This equivalence provides an important theoretical foundation for the performance analysis and optimization of massive MIMO systems.

Conventional time-domain coding schemes are limited in their ability to fully exploit the spatial freedom of massive MIMO. To overcome these deficiencies, spatiotemporal 2-D channel coding is proposed in \cite{you2022spatiotemporal}. In spatiotemporal coding schemes, the time-domain coding is extended to time and spatial domains. Channel coding is first performed in the time domain and then in the spatial domain across multiple data streams. The receiver improves system performance through joint decoding in the time and spatial domains, and low latency is ensured by decoding each codeword in parallel. However, despite its practical advantages, existing 2-D coding approaches are mainly empirical and lack a unified theoretical formulation and rigorous performance analysis framework.  Specifically, no general model exists to describe the structural equivalence and transformations among different 2-D coding constructions, limiting its design flexibility in massive MIMO systems. Moreover, the capacity-achieving capability of 2-D coding in spatiotemporal channels has not been theoretically studied, leaving a critical gap in validating its fundamental advantages. These limitations significantly hinder both the theoretical advancement and practical deployment of spatiotemporal coding schemes.


Polar codes, introduced by Arikan \cite{arikan2009channel}, have received much attention for their merit of achieving channel capacity on binary erasure channels (BEC) and binary symmetric channels (BSC). In the construction of polar codes, the key lies in the selection of information set. In traditional BSC channels, the reliability of each polarized channel is evaluated by recursively calculating its Bhattacharyya parameter and information bits are preferentially transmitted over highly reliable channels to approach the optimal performance with long codewords. However, in additive white Gaussian noise (AWGN) channels, the calculation of Bhattacharyya parameters is too complicated and lacks exact recursive formula \cite{arikan2009channel,korada2009polar}, which limits its practical application. To this end, Gaussian approximation (GA) is proposed to construct polar codes for AWGN channels by tracing the mean value of the polarized channel's log likelihood ratio (LLR) \cite{trifonov2012efficient,wu2014construction}. Although GA has already been widely used in the construction of polar codes for AWGN channels, the theoretical analysis is still incomplete regarding it capacity-achieving property, to the best of our knowledge. Moreover, in practical applications, polar codes require large code lengths to achieve a sufficient degree of channel polarization for approaching capacity, which consequently introduces considerable latency. Therefore, how to reduce the code length while ensuring the reliability has become an urgent problem in the research and application of polar codes.

Building upon these works, this paper seeks to incorporate the DoFs of spatiotemporal transmission to the inherent structure of polar codes. A new spatiotemporal 2-D polar coding scheme is proposed for massive MIMO with MMSE detection at receiver end, which performs polar transform  in both spatial and temporal domain. Most importantly, this work establishes the first unified theoretical framework for 2-D polar coding in massive MIMO systems, enabling systematic description, analysis, and design of spatiotemporal polar codes. To be specific, the deterministic approximation property of MMSE receivers allow us to model the massive spatial channels as multiple nearly-uniform Gaussian channels, and the traditional time-domain polar codes are extended to time-space domain by applying polar coding over multiple spatial streams. The Kronecker product of spatial and temporal component codes is equivalent to a polar code with the same length of the product code. By proportionally exchanging spatial DoF for temporal DoF, the time duration of a codeword can be reduced, thereby significantly reducing transmission latency. This is the specific demonstration of spatiotemporal exchangeability principle in the field of channel coding for massive MIMO communications \cite{you20236g}. Moreover, we theoretically prove that the proposed 2-D polar coding scheme retains the capacity-achieving property under short blocklength, while effectively reducing time-domain resource usage. This is of great importance to meeting the requirements of high reliability and low latency in the 6G eURLLC (enhanced URLLC) scenarios for massive MIMO systems \cite{you2021towards,rasti2022evolution}.

The main contributions of this paper are summarized as follows.

\begin{itemize}
	\item \textbf{Spatiotemporal polar coding theoretical framework:} We propose the first unified theoretical framework of spatiotemporal 2-D polar coding for massive MIMO systems, bridging the gap between empirical design and theoretical understanding. Guided by channel polarization theory and the principle of spatiotemporal exchangeability, the isomorphism property among different 2-D coding structures is formally characterized, enabling flexible and principled allocation of time and space resources of MIMO systems.
	
	\item \textbf{Theoretical proof of capacity-achieving property of spatiotemporal polar codes:} We theoretically analyze and prove that the proposed spatiotemporal 2-D polar codes can achieve the capacity of Gaussian channels even under short blocklength constraints and with large DoF. This is accomplished by leveraging the deterministic approximation theory of MMSE detection and deriving a BER upper bound using constructive GA analysis, thus establishing a solid theoretical foundation for their performance guarantee.
\end{itemize}

The rest of this article is organized as follows. Section II introduces the basic principle of traditional 1-D time-domain polar codes and the GA method. In Section III, the theoretical framework of spatiotemporal 2-D polar coding for massive MIMO is proposed. Section IV gives the capacity-achieving results of spatiotemporal polar codes based on GA. The simulation results are presented in Section V. Finally, Section VI concludes the article.

\section{Traditional 1-D polar codes and Gaussian approximation}\label{sec2}

\subsection{1-D Polar Codes}
Polar codes are constructed based on the phenomenon called channel polarization. Consider a binary input discrete memoryless channel $W:\mathcal{X}\rightarrow\mathcal{Y}$ with the input alphabet $\mathcal{X}=\{0,1\}$ and $\mathcal{Y}$ as the output. Through channel combining and splitting operations, the $N=2^{n}$ independent uses of the original channel $W$ can be transformed into a series of polarized channels $W_N^{(i)}:\mathcal{Y}\times\mathcal{X}^{i-1},i\in\{1,2,...,N\}$ with different reliabilities. In order to minimize the BER, the $K$ most reliable polarized channels are selected for transmitting information bits, and the remaining $N-K$ channels are used for transmitting frozen bits. Thus, an original sequence $u_1^N=(u_1,u_2,...,u_N)$ consisting of $K$ information bits and $N-K$ frozen bits is encoded into a codeword $x_1^N$ in the following manner.
\begin{equation}\label{eqn-1}
	x_1^N=u_1^NG_N,
\end{equation}
where $G_N=B_NF^{\otimes n}$ is the generator matrix, $B_N$ is the bit-reversal matrix, and $F^{\otimes n}$ denotes the $n$-th Kronecker product of the kernel matrix $F=\begin{bmatrix} 1&0\\1&1\end{bmatrix}$. The idea of channel polarization is embodied in the generator matrix $G_N$. The $N$ encoded bits $x_1^N=(x_1,x_2,...,x_N)$ are transmitted through $N$ independent channels $W$ with the same channel characteristics, which is equivalent to $u_1^N$ being transmitted through $N$ polarized channels $W_N^{(i)}$ with different channel characteristics. Arikan pointed out that when the code length $N$ approaches infinity as a power of 2, the channel will be fully polarized into two types of channels: completely reliable and completely unreliable, and the channel capacity can be reached by transmitting information bits on the completely reliable channels. The relevant results are given by Theorem 1 in reference [1].

\subsection{Gaussian Approximation}
Gaussian approximation is a construction method based on the simplification of probability density functions to reduce the complexity of the density evolution approach in designing polar codes. The key assumption is that the LLR follows a Gaussian distribution, and the reliability of each polarized channel can be calculated by recursively computing the mean of the LLR. 
We assume that BPSK modulation is used on an AWGN channel. Then we have
\begin{equation}\label{eqn-2}
	c_i=(1-2x_i),\quad1\leq i\leq N
\end{equation}
where $x_i$ is the $i$-th codeword and $c_i$ is the corresponding BPSK symbol. Then the received symbol is given by
\begin{equation}\label{eqn-3}
	y_i=c_i+n_i,\quad1\leq i\leq N 
\end{equation}
where $n_i\sim\mathcal{N}(0,\sigma^2)$ is the additive Gaussian white noise. Assuming that the original sequence is all zeros, then the BPSK sequence is all 1s, and the received symbol $y_i$ is a random variable following the Gaussian distribution $\mathcal{N}(1,\sigma^2)$. The LLR of the $i$-th bit can be calculated as
\begin{equation}\label{eqn-4}
	LLR_i=\ln\frac{P(y_i|0)}{P(y_i|1)}=\frac2{\sigma^2}y_i.
\end{equation} 
It can be inferred that $L_i{\sim}\mathcal{N}(\frac2{\sigma^2},\frac4{\sigma^2})$, i.e., the LLRs of the sub-channels in channel polarization process all follow a Gaussian distribution with a variance twice the mean. Denote the LLR mean of $W_N^{(i)}$ as $m_N^{(i)}$, and it can be recursively calculated by \cite{chung2001analysis}
\begin{align}
	\label{eqn-5} m_{2N}^{(2i-1)}&=\phi^{-1}\left(1-\phi\left(m_N^{(i)}\right)^2\right),\\
	\label{eqn-6} m_{2N}^{(2i)}&=2m_N^{(i)},
\end{align}
where $\phi(\gamma)=1-\int_{-\infty}^\infty\frac1{\sqrt{4\pi\gamma}}\tanh\left(\frac\alpha2\right)e^{-\frac{(\alpha-\gamma)^2}{4\gamma}}d\alpha$. [2] gives an approximation for $\phi(\gamma)$ as follows
\begin{equation}\label{eqn-7}
\phi (\gamma ) \approx \left\{ {\begin{array}{*{20}{c}}
		{{e^{a{\gamma ^c} + b}}}&,\ \ {\gamma  \le {\Gamma _{{\rm{th}}}}}\\
		{\sqrt {\frac{\pi }{\gamma }} {{\rm{e}}^{ - \frac{\gamma }{4}}}\left( {1 - \frac{{10}}{{7\gamma }}} \right)}&,\ \ {\gamma  > {\Gamma _{{\rm{th}}}}}
\end{array}} \right.
\end{equation}
where $(a,b,c)$ is set as $(-0.4527, 0.0218, 0.86)$, and the threshold is $\Gamma_{\mathrm{th}}\approx10$.

\begin{figure*}[t]
	\centering
	\includegraphics[width=6.5in]{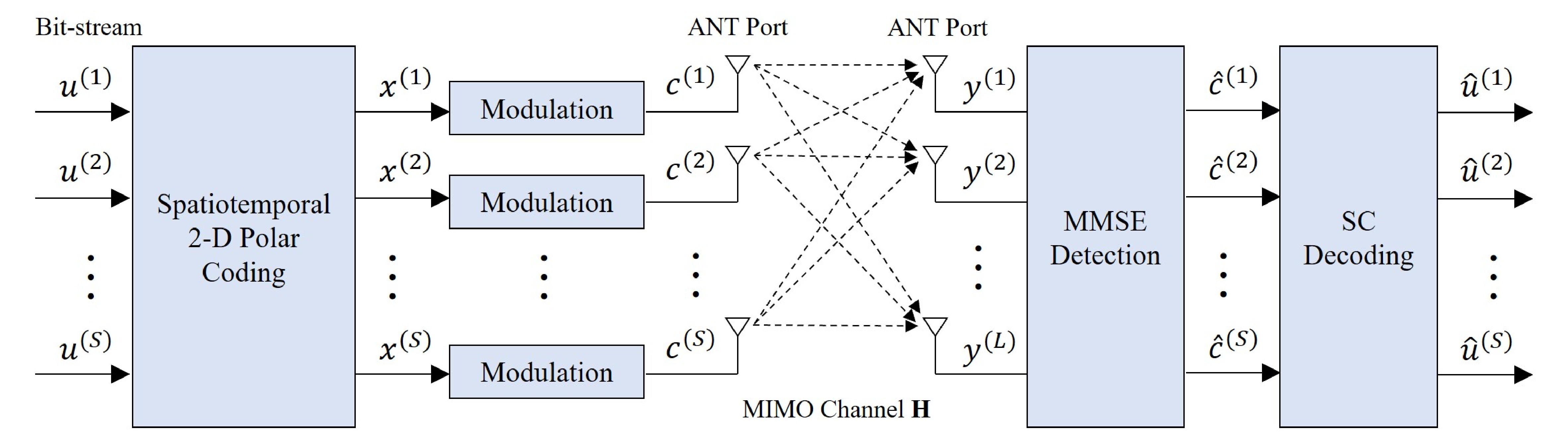}
	\caption{The transmitting system architecture with spatiotemporal 2-D polar coding}
	\label{fig1}
\end{figure*}

\section{Proposed Spatiotemporal 2-D polar coding}\label{sec3}
\subsection{Channel Model}
Consider a MIMO channel $\mathbf{H}$ with $S$ transmit antennas and $L$ receive antennas, operating over quasi-static flat fading channels, where the random fading coefficients remain constant over the duration of each codeword. The relationship between the channel input and output can be expressed as:
\begin{equation}
	{\mathbf{Y}=\mathbf{H}\mathbf{X}+\mathbf{Z}},
\end{equation}
where $\mathbf{X}\in\mathbb{C}^{S\times N}$ represents the signal transmitted over $N$ time samples, and each column of $\mathbf{X}$ is normalized to have unit norm. $\mathbf{Y}\in\mathbb{C}^{L\times N}$ is the corresponding received signal. The channel matrix is $\mathbf{H}\in\mathbb{C}^{L\times S}$ which contains random complex fading elements. Each element is an i.i.d. $\mathcal{CN}(0,1)$ complex Gaussian random variable, which remains constant over the $N$ time samples. Let $\gamma = \frac{S}{L}$ indicate the ratio of transmitting antenna to receiving antenna. Assuming that the transmitter has unknown channel state information (CSI) due to the insufficient time for the receiver to feedback CSI in a URLLC transmission, equal power allocation is adopted across transmit antennas. $\mathbf{Z}\in\mathbb{C}^{L\times N}$ is the additive noise signal at the receiver, which is independent of $\mathbf{H}$ and each of its elements follows a $\mathcal{CN}(0,1)$ complex Gaussian distribution.

When MMSE detection is employed, the SINR on the $k$-th spatial stream can be expressed as \cite{poor1997probability,tse2000linear,verdu1998multiuser,paulraj2003introduction}
\begin{equation}
	\mathrm{SINR}_k=\frac{1}{\mathrm{MMSE}_k}-1=\frac{1}{\left[\left(\mathbf{I}_{S}+\frac{1}{S\sigma^2}\mathbf{H}^\dagger\mathbf{H}\right)^{-1}\right]_{kk}}-1,
\end{equation}
where $\left(\ \right)^\dagger$ denotes the Hermitian transpose and $\mathbf{I}_{S}$ is a $S \times S$ identity matrix. Based on the random matrix theory, the $\mathrm{SINR}_k$ asymptotically follows a normal distribution when the antenna dimension tends to infinity, as depicted in the following\cite{li2005distribution}.
\begin{equation}
	\label{normal_dist}
	\begin{aligned}
		\mathrm{SINR}_k\sim \mathcal{N}\big(\tfrac{L-S}{L}\Sigma_k+\tfrac{S}{L}\Sigma_k\mu_{\gamma},
		\tfrac{L-S}{L^2}\Sigma_k^2+\tfrac{S}{L^2}\Sigma_k^2\sigma_{\gamma}^2\big),
	\end{aligned}
\end{equation}
where $\Sigma_k$, $\mu_{\gamma}$ and $\sigma_{\gamma}^2$ are parameters determined by $\gamma$ and can be calculated according to reference \cite[(23)(24)]{li2005distribution}. Note that the variance of $\mathrm{SINR}_k$ diminishes to $0$ at a rate of $\frac{1}{L}$, indicating that the spatial channels become asymptotically deterministic for large antenna array. Furthermore, the parameter $\Sigma_k$, which determines the SINR of the $k$-th spatial stream, is approximately constant across the spatial streams in scenarios with equicorrelated and uncorrelated transmitter antennas. Such spatial uniformity is also demonstrated for a variety of realistic channels models.

This deterministic approximation property suggests that the massive MIMO channels with MMSE receivers can be approximately equivalent to  parallel Gaussian channels. Furthermore, spatial uniformity allows channel polarization across these streams, in a manner analogous to polarization in the time domain. Therefore, a 2-D spatiotemporal polar encoding scheme is proposed, incorporating polar transform both in temporal and spatial domain. The transmitting system architecture is illustrated in Figure \ref{fig1}, and the specific implementation of the module ``Spatiotemporal 2-D polar coding" in given in Figure \ref{fig2}, which will be introduced in detail as below in Part B.

\subsection{Spatiotemporal Polar Coding Theoretical Framework}
The essence of time-domain polar codes is the $n$-stage polarization process, which requires long codeword to achieve the ideal polarization effect. Inspired by the spatiotemporal exchangeability principle, 1-D time-domain polarization can be extended to the spatial-temporal domain. This extension is achieved by leveraging a large number of spatial streams, effectively trading spatial DoF for temporal DoF. Consequently, the code length can be significantly reduced while maintaining the same depth of polarization. In the following, the theoretical framework of spatiotemporal 2-D polar codes is constructed and further elaborated with examples.

Similar to the 5G NR polar codes standard developed by 3GPP, we remove $B_N$ and a new generator matrix $F_N=F^{\otimes n}$ is used in this paper. The codeword after time-domain polar coding can then be rewritten as
\begin{equation}\label{eqn-8}
	x_1^N=u_1^NF_N.
\end{equation}
In the proposed spatiotemporal 2-D polar coding framework, we define the generator matrices of spatial component and time-domain component as $F_S$ and $F_T$, respectively. $S=2^s$ is the number of bit streams in spatial domain, $T=2^t$ is the blocklength in time domain, $N=S\times T=2^n$ represents the total coding length, and $n=s+t$. When $S=1$($s=0$), we have $N=T$, and the 2-D coding is reduced to 1-D time-domain coding; Similarly, when $T=1$($t=0$), we have $N=S$, and the 2-D coding is reduced to 1-D spatial coding. Note that $T$ and $S$ cannot be 1 at the same time. According to the rules of Kronecker product, $F^{\otimes n}=F^{\otimes (s+t)}=F^{\otimes s}\otimes F^{\otimes t}$, i.e., $F_N=F_S\otimes F_T$. The spatiotemporal 2-D polar coding process is explained in detail below.

Divide the original bit sequence $u_1^N$ into $S$ segments, each length $T$, and reshape it into a two-dimensional information matrix form as follows:
\begin{equation}
	u_1^N=U=[u_{ij}]_{i=1,...,S,j=1,...,T}.
\end{equation}
Firstly, \textbf{time-domain polar coding} with length $T$ is performed on each row $u^{(i)}=[u_{i1},u_{i2},...,u_{iT}]$ of $U$ as
\begin{equation}
	v^{(i)}=u^{(i)}F_T,
\end{equation}
and the time-domain encoding matrix is obtained as
\begin{equation}
	V=\begin{bmatrix}v^{(1)}\\v^{(2)}\\\vdots\\v^{(S)}\end{bmatrix}=[v_1,v_2,...,v_T].
\end{equation}
Then, \textbf{spatial-domain polar coding} with length $S$ is performed on each column $v_j$ of $V$ as
\begin{equation}
	x_j=F_Sv_j,
\end{equation}
and the final spatiotemporal codeword matrix is obtained as
\begin{equation}
	X=[x_1,x_2,...,x_T].
\end{equation}
Thus, the joint spatiotemporal 2-D polar encoding is completed. The encoding process can be described as a linear mapping that $$\mathrm{Enc}: \mathbb{F}_2^{S\times T} \to \mathbb{F}_2^{S\times T}: (\cdot)  \mapsto F_S (\cdot) F_T, $$where $ \mathbb{F}_2$ is the binary field. Compared with traditional time-domain polar codes, the spatiotemporal polar codes involve parallel processing of multiple data streams, and the polarized channels are no longer $n$-times multiplexing of the same physical channel, but the result of the polarization of both time domain and spatial domain across several data streams.

Next, it can be shown that the spatiotemporal encoding with a spatial length of $S$ and a temporal length of $T$ is equivalent to 1-D polar codes of length $N=S\times T$. The information matrices $U,\ V$ and the 2-D encoding matrix $X$ can be reformulated into 1-D row vectors as follows.
\begin{equation}
	U^{\prime}=\begin{bmatrix}u^{(1)},u^{(2)},...,u^{(S)}\end{bmatrix},
\end{equation}
\begin{equation}
	V^{\prime}=\left[v^{(1)},v^{(2)},...,v^{(S)}\right]=U^{\prime}(I_S\otimes F_T),
\end{equation}
\begin{equation}
	X^{\prime}=\begin{bmatrix}v^{(1)},v^{(2)},...,v^{(S)}\end{bmatrix}=V'(F_S\otimes I_T),
\end{equation}
where $I_N$ denotes the $n$-th order identity matrix. By the mixed-product property of the Kronecker product, $(\mathrm{A}\otimes\mathrm{B})(\mathrm{C}\otimes\mathrm{D})=\mathrm{AC}\otimes\mathrm{BD}$. Thus we have
\begin{equation}
	\begin{aligned}
		X^{\prime}&=U^{\prime}(I_S\otimes F_T)(F_S\otimes I_T)\\
		&=U^{\prime}(F_S\otimes F_T)\\
		&=U^{\prime}F_N,
		\end{aligned}
\end{equation}
where the generator matrix of 1-D polar code is the Kronecker product of generator matrices of the spatial and temporal components. 
Therefore, the proposed spatiotemporal 2-D polar coding can achieve the same encoding results as 1-D polar code, provided that their code dimensions are equal. In addition, the 1-D polarization process can be decomposed into spatial and temporal polarization sub-processes, as is given by Lemma \ref*{le1} in Section \ref{sec4}.

\begin{figure}[!t]
	\centering
	\includegraphics[width=3.2in]{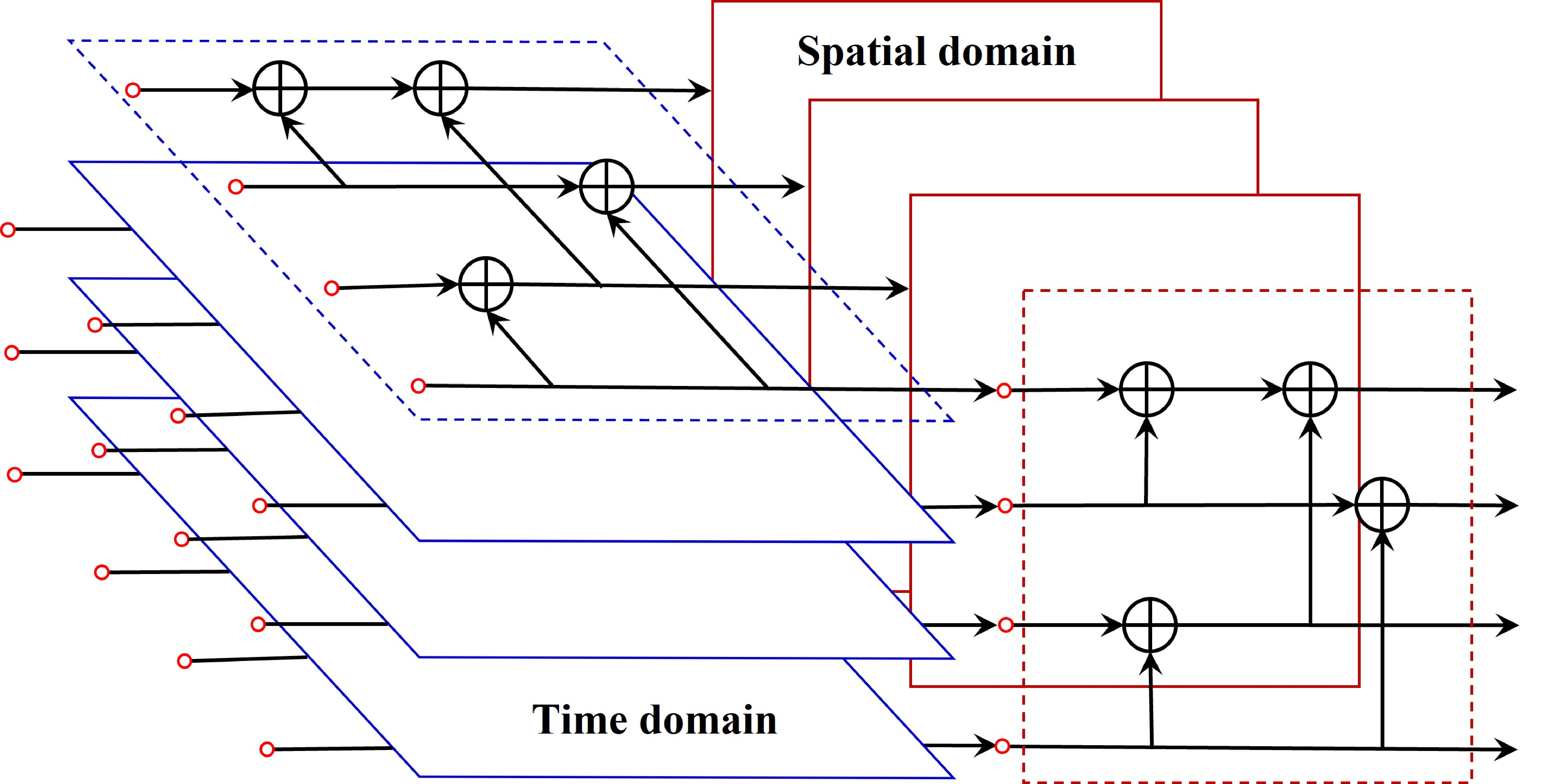}
	\caption{Spatiotemporal 2-D polar coding structure}
	\label{fig2}
\end{figure}

The encoding structure of the proposed spatiotemporal 2-D polar codes is shown in Figure \ref{fig2}. The coding process above demonstrates that 1-D polar codes in the time domain can be transformed into spatiotemporal 2-D polar codes by performing polar coding in the time and space domains respectively. This extension allows for achieving performance comparable to 1-D time-domain codes with significantly shorter temporal blocklengths. In addition, the coding mode above is called \textbf{time-space mode}, which means that polar codes are fist performed in the time domain and subsequently in the spatial domain across several data streams. It is noted that we can also perform the \textbf{space-time mode}, which means that the order of time-domain coding and spatial coding can be exchanged while the coding result will not be affected. One only needs to adjust the operation order of the generator matrices $F_S$ and $F_T$ accordingly.

\textbf{Examples:} Specific examples are given below to elaborate the proposed spatiotemporal 2-D polar coding scheme.

\begin{figure}[!t]
	\centering
	\subfigure[Time-domain polar coding structure for $N=4$]{
		\label{fig3_a}
		\includegraphics[width=3in]{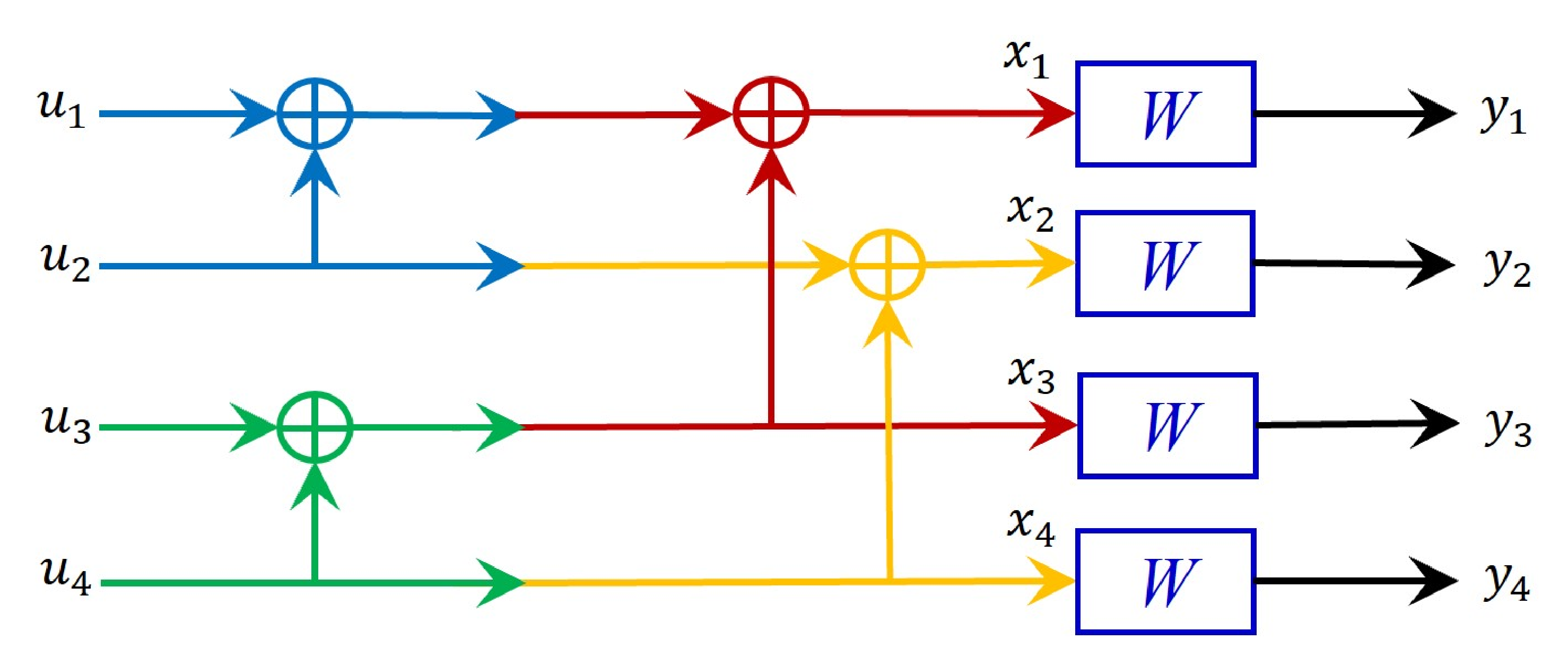}}
	\quad\\
	\subfigure[Spatiotemporal polar coding structure for $S=2$ and $T=2$]{
		\label{fig3_b}
		\includegraphics[width=3.2in]{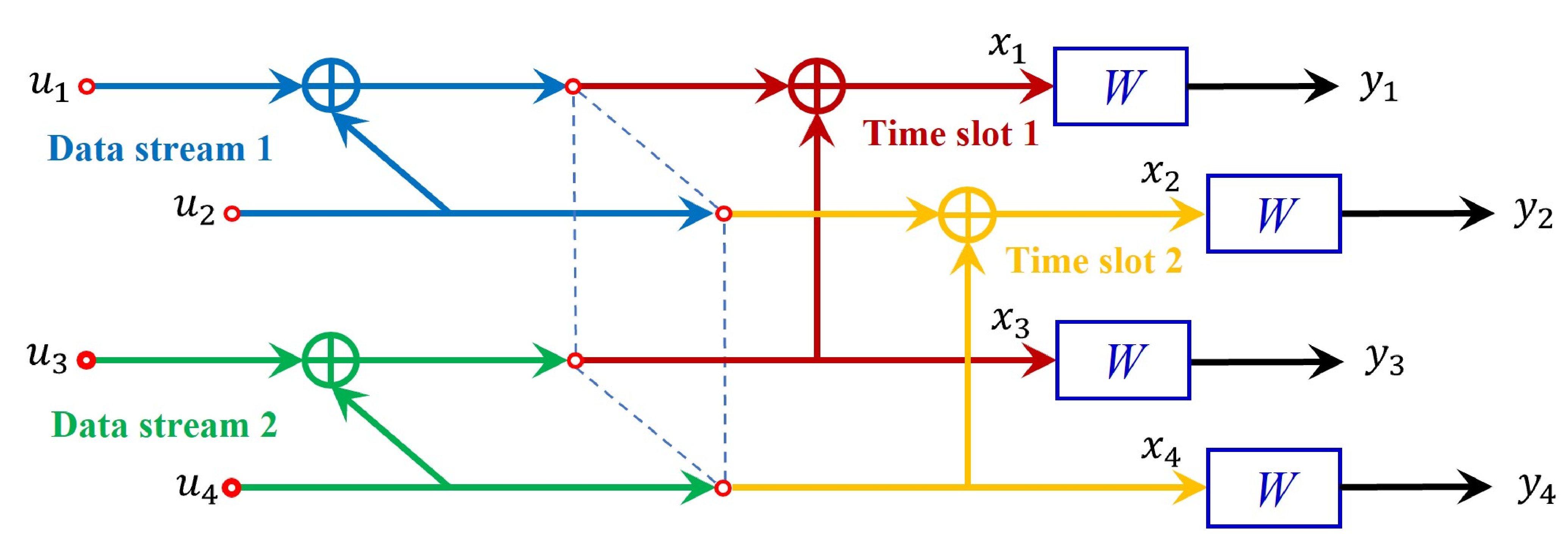}}
	\caption{1-D and 2-D polar coding structures for $N=4$}
	\label{fig3}
\end{figure}

Consider the 1-D time-domain polar codes with $N=4$, the coding sequence is given in \eqref{eqn-10} and its time-domain coding structure is shown in Figure \ref{fig3_a}.
\begin{equation}\label{eqn-10}
	\begin{aligned}
	u_{1}^{4}F_{4}&=[u_{1},u_{2},u_{3},u_{4}]F_{4}\\
	&=[u_{1}\oplus u_{2}\oplus u_{3}\oplus u_{4},u_{2}\oplus u_{4},u_{3}\oplus u_{4},u_{4}]\\
	&=[x_{1},x_{2},x_{3},x_{4}].
	\end{aligned}
\end{equation}

When adopting the spatiotemporal polar coding with $S=2$ and $T=2$, the 2-D information matrix is $U=[u_1^2;u_3^4]$. After time-domain polar coding, the encoding matrix is
\begin{equation}
	V=\begin{bmatrix}u_{1}^{2}F_{2}\\u_{3}^{4}F_{2}\end{bmatrix}=\begin{bmatrix}u_{1}\oplus u_{2},u_{2}\\u_{3}\oplus u_{4},u_{4}\end{bmatrix}=\begin{bmatrix}v_{1},v_{2}\end{bmatrix}.
\end{equation}
After spatial-domain polar coding, the spatiotemporal polar encoding matrix is
\begin{equation}\label{eqn-11}
	\begin{aligned}
		X&=\begin{bmatrix}v_1'F_2\\v_2'F_2\end{bmatrix}=\setlength{\arraycolsep}{1pt}\begin{bmatrix}u_1\oplus u_2\oplus u_3\oplus u_4,&u_3\oplus u_4\\u_2\oplus u_4,&u_4\end{bmatrix}\\
		&=\begin{bmatrix}x_1,x_3;x_2,x_4\end{bmatrix}.
	\end{aligned}
\end{equation}

It can be seen that by polar coding from both time and spatial domains, the same results as 1-D time-domain polar coding can be obtained and its spatiotemporal coding structure is shown in Figure \ref{fig3_b}.

\begin{figure}[!t]
	\centering
	\subfigure[Time-domain polar coding structure for $N=8$]{
		\label{fig4_a}
		\includegraphics[width=3.2in]{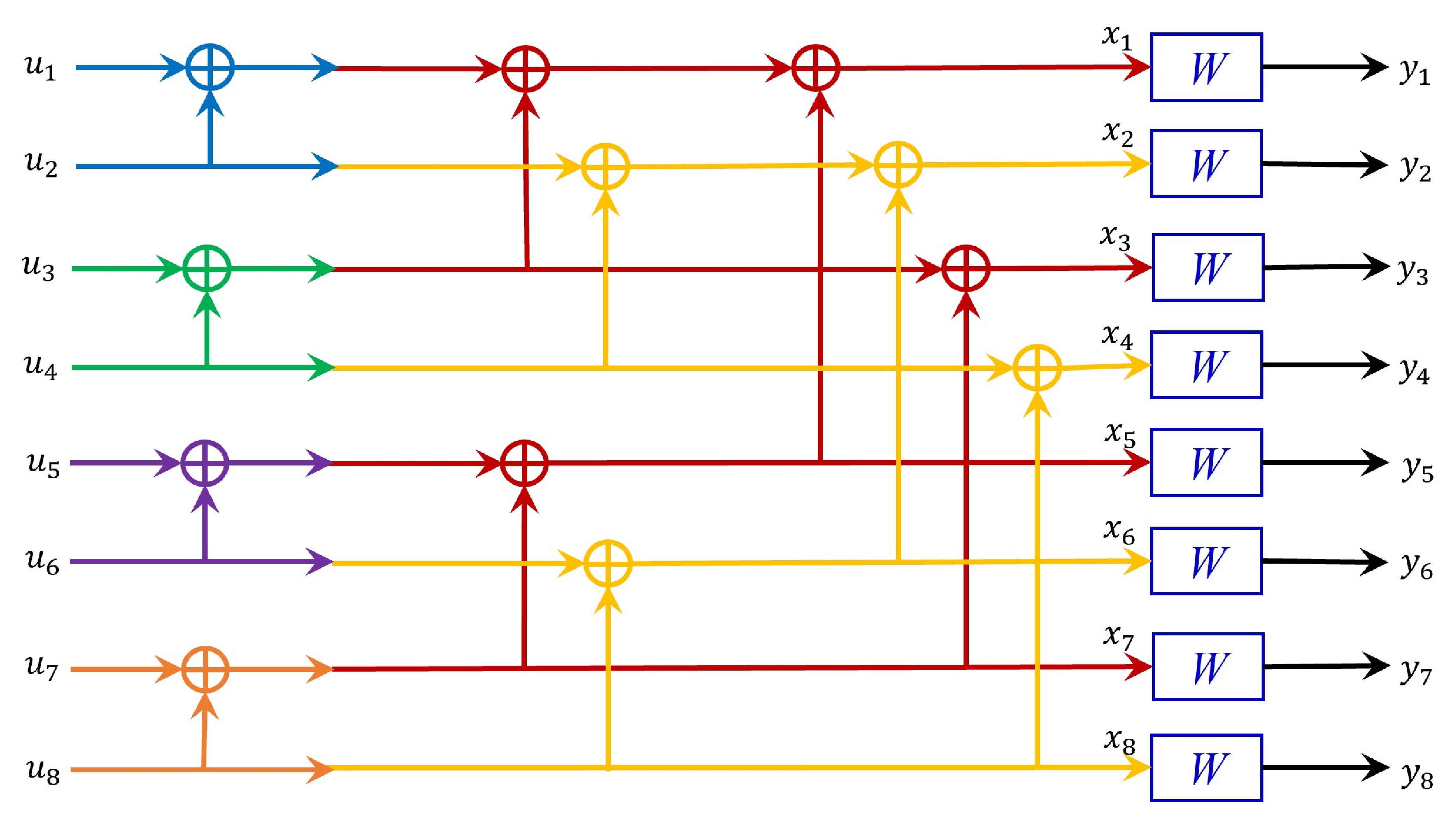}}
	\quad\\
	\subfigure[Spatiotemporal polar coding structure for $S=2$ and $T=4$]{
		\label{fig4_b}
		\includegraphics[width=3.2in]{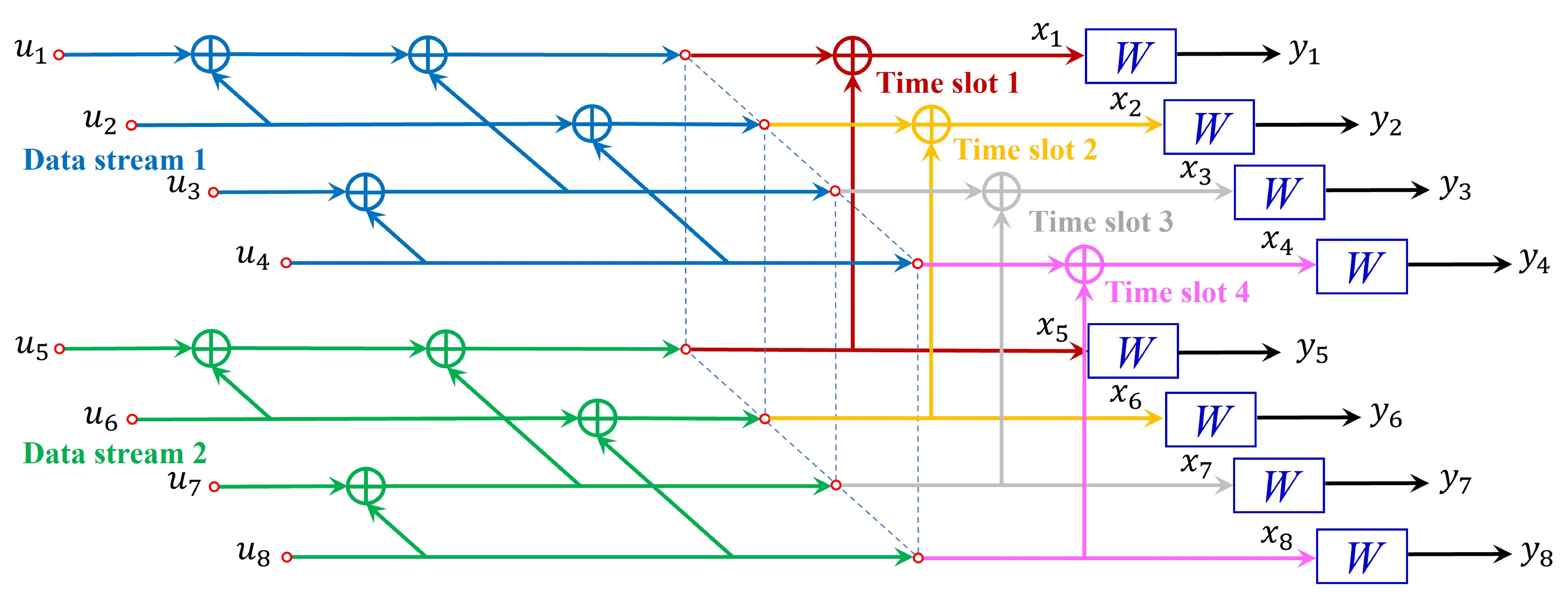}}
	\quad\\
	\subfigure[Spatiotemporal polar coding structure for $S=4$ and $T=2$]{
		\label{fig4_c}
		\includegraphics[width=3.2in]{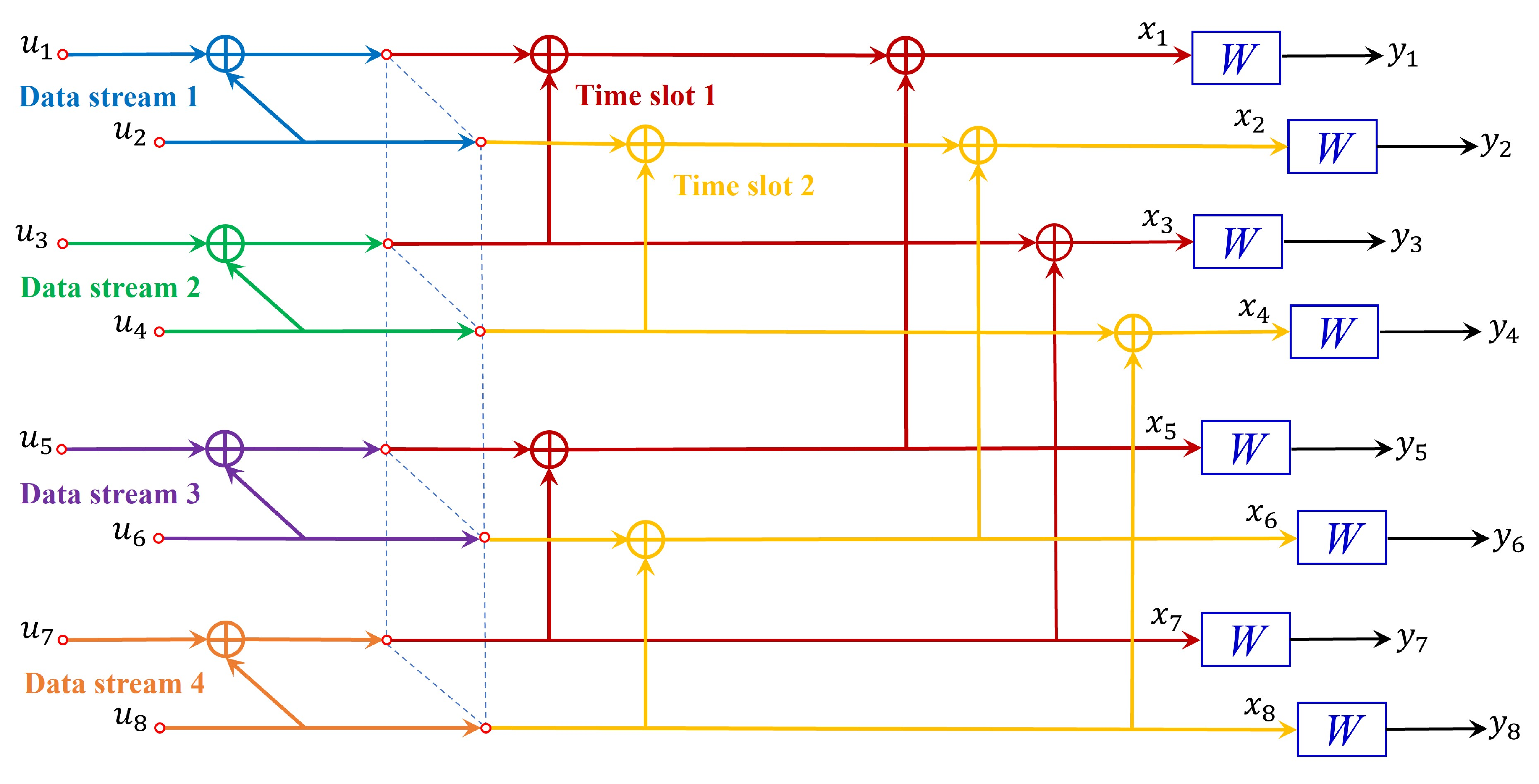}}
	\caption{1-D and 2-D polar coding structures for $N=8$}
	\label{fig4}
\end{figure}

Now consider a polar coding structure with $N=8$. In contrast to the previous example, the spatiotemporal polar coding with $N=8$ has different combinations of $S$ and $T$. Firstly, its 1-D coding sequence is given in \eqref{eqn-12} and the corresponding structure in given in Figure \ref{fig4_a}.
\begin{equation}\label{eqn-12}
	\begin{aligned}
		u_1^8 F_8&=[u_{1},u_{2},u_{3},u_{4},u_{5},u_{6},u_{7},u_{8}]F_{8}\\
		&=[x_{1},x_{2},x_{3},x_{4},x_{5},x_{6},x_{7},x_{8}].
	\end{aligned}
\end{equation}

When adopting the spatiotemporal polar coding, there exist four different combinations of $S$ and $T$:
\begin{itemize}
	\item $S=1,T=8$: In this case, the 2-D coding is simplified to 1-D time-domain coding.
	\vspace{2pt}
	\item $S=2,T=4$: The 2-D information matrix is $U=[u_1^4;u_5^8]$, and the encoding process is as follows.
	
	Time-domain polar coding:
	\vspace{-1pt}
	\begin{equation}\label{eqn-13}
	\begin{aligned}
		V&=\begin{bmatrix}u_1^4F_4\\u_5^8F_4\end{bmatrix}=\begin{bmatrix}u_1\oplus u_2\oplus u_3\oplus u_4,u_2\oplus u_4,u_3\oplus u_4,u_4\\u_5\oplus u_6\oplus u_7\oplus u_8,u_6\oplus u_8,u_7\oplus u_8,u_8\end{bmatrix}\\
		&=[v_1,v_2,v_3,v_4]
		\end{aligned}
	\end{equation}
	
	\vspace{-2pt}
	Spatial-domain polar coding:
\vspace{-1pt}
	\begin{equation}\label{eqn-13}
	\begin{aligned}
		X&=\begin{bmatrix}v_1'F_2;v_2'F_2;v_3'F_2;v_4'F_2\end{bmatrix}\\
		&=\begin{bmatrix}x_1,x_5;x_2,x_6;x_3,x_7;x_4,x_8\end{bmatrix}
	\end{aligned}
	\end{equation}

	The coding structure is shown in Figure \ref{fig4_b}.
	\vspace{2pt}
	\item $S=4,T=2$: The 2-D information matrix is $U=[u_1^2;u_3^4;u_5^6;u_7^8]$, and the encoding process is as follows.
	
	Time-domain polar coding:
	\begin{equation}\label{eqn-13}
		V=\begin{bmatrix}u_1^2F_2\\u_3^4F_2\\u_5^6F_2\\u_7^8F_2\end{bmatrix}=\begin{bmatrix}u_1\oplus u_2,u_2\\u_3\oplus u_4,u_4\\u_5\oplus u_6,u_6\\u_7\oplus u_8,u_8\end{bmatrix}=[v_1,v_2]
	\end{equation}
	Spatial-domain polar coding:
		\begin{equation}\label{eqn-13}
		\begin{aligned}
			X&=\begin{bmatrix}v_1'F_4;v_2'F_4\end{bmatrix}\\
			&=\begin{bmatrix}x_1,x_3,x_5,x_7;x_2,x_4,x_6,x_8\end{bmatrix}
		\end{aligned}
	\end{equation}
	
	The coding structure is shown in Figure \ref{fig4_c}.
	\vspace{2pt}
	\item $S=8,T=1$: In this case, the 2-D coding is simplified to 1-D spatial coding, which has the same structure as Figure \ref{fig3_a}.
\end{itemize}

Through the analysis of the above examples, we draw the following conclusions: within the framework of the spatiotemporal 2-D polar coding scheme, equivalent coding structures can be achieved by appropriately configuring time-domain ($T$) and spatial-domain ($S$) parameters, which is a specific application of the spatiotemporal exchangeability theory in the field of channel coding. Specifically, by increasing the scale of the transceiver antenna array (i.e., expanding the spatial domain), the spatial DoF can be effectively utilized to substitute for the time DoF, thereby shortening the blocklength and effectively reducing latency. This characteristic holds significant theoretical importance and practical value to meet the requirements of high reliability and low latency in 6G URLLC scenarios. In the next section, a detailed analysis will be conducted to demonstrate the spatiotemporal polar codes' ability in approaching the channel capacity under GA method.

\section{capacity-achieving characteristics of spatiotemporal polar codes}\label{sec4}

Channel polarization is one of the crucial characteristics of polar codes. Spatiotemporal 2-D polar codes, as the extension of traditional 1-D polar codes in the time and spatial domains, also possess the key attributes of channel polarization. In this section, we extend the 1-D time-domain channel polarization theory in reference [1] to 2-D scenarios, consider spatiotemporal polar coding on multiple parallel Gaussian channels, and prove that when the product of $S$ and $T$ tends to infinity, the spatiotemporal polar codes also exhibit a polarization phenomenon similar to that of 1-D polar codes: the capacity of some polarized channels approaches 1, while that of the others approaches 0. Meanwhile, considering that the Bhattacharyya parameter is a non-constructive coding method, it is difficult to make accurate recursive polarization calculations in Gaussian channels. Therefore, the capacity-achieving property of the constructed polarization codes based on GA is further proved in this section. 

It is immediate from the equivalence of generator matrix that any two 2-D polar codes of the same code dimensions are functionally equivalent. Nevertheless, the method for decomposing 2-D polarization in the spatial and temporal domains still requires further investigation. Under the assumption of spatial channel consistency, we consider multiple spatial parallel binary memoryless symmetric AWGN channels $W$ with capacity $C$. Assuming that spatiotemporal 2-D polar coding is adopted, let $T=2^t$ be the time-domain blocklength and $S=2^s$ be the spatial codeword length, then the spatiotemporal polarization sub-channels can be represented by two-dimensional variables of time and space:
	\begin{equation}\label{eqn-15}
		\{W_{TS}^{(i,j)}\colon1\leq i\leq T,1\leq j\leq S\}.
	\end{equation}

	Firstly, the polarized subchannel $W_{2^{s+t}}^{(i,j)}$ can be represented as $W_{p_1p_2...p_sq_1q_2...q_t}$, such that the following two equations hold.
	\begin{align}\label{eqn-16}
		i&=1+\sum_{k=1}^tp_k2^{s-k},\\
		j&=1+\sum_{k=1}^sq_k2^{t-k}.
	\end{align}
	The index $p_1p_2...p_tq_1q_2...q_s$ of the polarized subchannel $W_{p_1p_2...p_tq_1q_2...q_s}$ can be viewed as being generated by a sequence of random variables $P_1,P_2,...,P_t,Q_1,Q_2,...,Q_s$, where each random variable $P_i$ (or $Q_j$) is a Bernoulli random variable. Define a stochastic process $\{K_{t,s}:t\geq0,s\geq0\}$, and consider an instance of this stochastic process as $W_{p_1p_2...p_tq_1q_2...q_s}$. Considering the polarization process of spatiotemporal 2-D polar codes, each polarization split can be performed based on either the time domain or the spatial domain. If the polarization is performed based on the time domain, resulting in $K_{t+1,s}$, its value can be either $K_{t+1,s}=W_{{p_{1}}p_{2}...p_{t}0q_{1}q_{2}...q_{s}}$ or $K_{t+1,s}=W_{{p_{1}}p_{2}...p_{t}1q_{1}q_{2}...q_{s}}$, each occurring with a probability of $1/2$. Similarly, if the polarization is performed based on the spatial domain, resulting in $K_{t,s+1}$, its value can be either $K_{t,s+1}=W_{p_1p_2...p_tq_1q_2...q_s0}$ or $K_{t,s+1}=W_{p_1p_2...p_tq_1q_2...q_s1}$, also each occurring with a probability of $1/2$. The polarization processes based on the time domain and the spatial domain can be respectively expressed as:
	\begin{itemize}
		\item time-domain polarization: 
		\vspace{3pt}
		
		$(W_{T\cdot S}^{(i,j)},W_{T\cdot S}^{(i,j)})\to(W_{2T\cdot S}^{(2i-1,j)},W_{2T\cdot S}^{(2i,j)})$
		\vspace{3pt}
		\item spatial domain polarization:
		\vspace{3pt}
		
		$(W_{T\cdot S}^{(i,j)},W_{T\cdot S}^{(i,j)})\to(W_{T\cdot2S}^{(i,2j-1)},W_{T\cdot2S}^{(i,2j)})$
		\vspace{2pt}
	\end{itemize}
	
	In spatiotemporal 2-D polar coding, the different choices of performing polarization in either the time domain or the spatial domain fundamentally lead to distinct spatiotemporal coding structures. For example, in 1-D time-domain coding, a polar code with $N=4$ is further polarized to yield a polar code with $N=8$, as shown from Figure \ref{fig3_a} to \ref{fig4_a}. In contrast, for a spatiotemporal 2-D polar code with dimensions $T\times S=2\times2$, as shown in Figure \ref{fig3_b}, if polarization is performed in the time domain, a polar code of dimensions $T\times S=4\times2$ is obtained; if polarization is performed in the spatial domain, a polar code of dimensions $T\times S=2\times4$ is obtained. These two cases correspond to the coding structures illustrated in Figure \ref{fig4_b} and \ref{fig4_c}, respectively.
	
	As described in the previous section regarding the encoding process of spatiotemporal polar codes, the 2-D polar encoder is equivalent to a 1-D polar encoder of the same code dimension $N$. When the total encoding length $N$ is given, any two different 2-D polar coding structure are considered to be equivalent. Furthermore, for spatiotemporal channel with almost uniform SINR, it can be shown through induction that they have equivalent polarization processes as described in the following Lemma.
	\begin{lemma}\label{le1}
		For the same polarized subchannels, the results of further polarization based on either the time domain or the spatial domain are equivalent.
		\begin{IEEEproof}
			For the subchannel $W_{p_0p_1\ldots p_{t-1}q_0q_1\ldots q_{s-1}}$, the time-domain and spatial-domain parts of the subscript are denoted as $p_{\mathrm{old}}=[p_{0}p_{1}\ldots p_{t-1}]$ and $q_{\mathrm{old}}=[q_{0}q_{1}\ldots q_{s-1}]$, respectively. Let $b_{\mathrm{old}}=[p_{\mathrm{old}},q_{\mathrm{old}}]=[b_0b_1\ldots b_{n-1}]$. Now consider its further polarization, i.e., the $(s+t+1)$-{th} step in the polarization process. If polarization is based on the spatial domain, the polarized subchannel $W_{p_0p_1\ldots p_{t-1}q_0q_1\ldots q_{s-1}q_s}$ is obtained, with subscripts $p_\mathrm{new}^s=[p_\mathrm{old}]$ and $q_\mathrm{new}^s=[q_\mathrm{old},q_s]$. Let $b_{\mathrm{new}}^{s}=[p_{\mathrm{new}}^{s},q_{\mathrm{new}}^{s}]=[b_{0}b_{1}\ldots b_{n-1}b_{n}]$. If polarization is based on the time domain, the polarized subchannel $W_{p_0p_1\ldots p_{t-1}p_tq_0q_1\ldots q_{s-1}}$ is obtained, with subscripts $p_{\mathrm{new}}^t=[p_{\mathrm{old}},p_t]$ and $q_{\mathrm{new}}^t=[q_\mathrm{old}]$. Then, $b_{\mathrm{new}}^{t}=[p_{\mathrm{new}}^{t},q_{\mathrm{new}}^{t}]=[b_{0}b_{1}\ldots b_{t-1}b_{n}b_{t}\ldots b_{n-1}]$. On the other hand, each codeword in polar codes can be represented as a binary tree, where each node represents a bit, and the path from the root node to a leaf node represents an encoded bit sequence. $b_{\mathrm{new}}^{s}$ and $b_{\mathrm{new}}^{t}$ are two different permutations of the same set $\{b_{0},b_{1},\ldots,b_{n-1},b_{n}\}$. Their corresponding permutations belong to the same permutation group and can be represented by two binary trees with the same leaf nodes but different structures. According to the encoding principle of polar codes, the bit $u_i$ generates the codeword $x_i$ by associating ``0" with the ``XOR" operation and ``1" with the ``dot" product operation in the binary expansion of the integer $i$. The different binary tree structures of the two polar codes mentioned above only change the order of the ``XOR"  and ``dot" product operations. Therefore, the two binary trees are equivalent under a certain traversal, and the same decoder can be used to perform decoding on both permutations \cite{doan2018decoding,korada2009polar}.
			
			Figure \ref{fig5} shows the two different binary tree structures of the polar codes with $N=4$ after polarization based on the time domain and the spatial domain. It can be seen that both structures yield the same polarization encoding results.
			\begin{figure}[!t]
				\centering
				\includegraphics[width=3.2in]{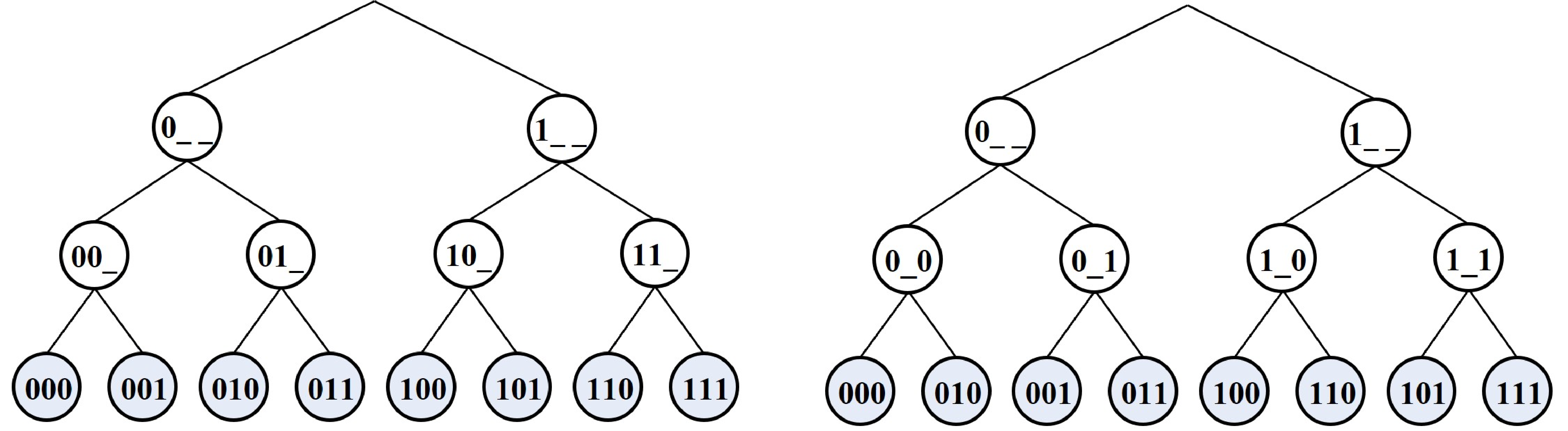}
				\caption{Different binary tree structures of spatiotemporal polar codes with $N=S\times T=8$}
				\label{fig5}
			\end{figure}
			
		\end{IEEEproof}
	\end{lemma}
	
	From Lemma \ref{le1}, it follows that the results of polarization based on the time domain or the spatial domain are equivalent for the same polarized subchannel. Therefore, different 2-D coding structures are isomorphic, differing only in the specific positions where polarization occurs. This fully demonstrates the equivalence of different permuted factor graphs for polar codes, where the bit-reversal permutation matrix $B_N$ mentioned above is just a special case. Such equivalent is a direct result from the automorphism groups of polar codes \cite{geiselhart2021automorphism}, making them highly suitable for adopting a spatiotemporal 2-D coding structure to achieve flexible allocation of time and space resources. The time-domain blocklength $T$ and the spatial-domain code length $S$ can be arbitrarily combined, as long as the condition $N=T\times S$ is satisfied. Consequently, the analysis of spatiotemporal 2-D polar codes can be simplified to the study of their most basic form, namely, 1-D polar codes.
	
	Since a binary-input memoryless symmetric AWGN channel can be discretized into a discrete memoryless channel (DMC), the conclusions regarding channel polarization results on B-DMC channels directly hold \cite[Theorem 1]{arikan2009channel}: For any fixed $\delta\in(0,1)$, as $TS$ goes to infinity, some polarized channels have a capacity approaching 1, i.e., $I(W_{TS}^{(i,j)})\in(1-\delta,1]$, while that of the others approaches 0, i.e., $I(W_{TS}^{(i,j)})\in[0,\delta)$. The number of polarized channels with capacity approaching 1 tends to $TSC$, and the number of polarized channels with capacity approaching 0 tends to $TS(1-C)$.

The conclusion above presents the polarization results for spatiotemporal 2-D polar coding under the assumption of consistent spatial channels. The analysis is also applicable to polarization in spatial channels with non-uniform SINRs, which is currently under active investigation in our research. Furthermore, the extent of channel polarization achieved when using the GA construction method is provided by the following Theorem \ref{th1}.

\begin{theorem}\label{th1}
	Consider multiple spatial parallel binary memoryless symmetric AWGN channels $W$ with capacity $C$. Assuming that spatiotemporal 2-D polar coding is adopted, for sufficiently large  $T\times S$ and any fixed code rate $R<C$, there exists an ordered pair set $\mathcal{A}_{TS}\subset\{(1,1),\cdots(i,j),\cdots,(T,S)\}$ such that $|\mathcal{A}_{TS}|\geq TSR$ and the BER $P(W_{TS}^{(i,j)})=Q\left(\sqrt{m_{TS}^{(i,j)}/2}\right)<O(N^{-\frac{2.33}{2}})$ holds for any $(i,j)\in\mathcal{A}_{TS}$. $m_{TS}^{(i,j)}$ is the LLR mean of the polarized channel. That is to say, there exist some spatiotemporal polarized subchannels whose channel capacities approach 1.
\end{theorem}

\begin{IEEEproof}
	We consider the most basic form of spatiotemporal polar codes. In this case, $S=1(s=0), N=T,W_{TS}^{(i,j)}=W_N^{(i,1)}$, which is abbreviated as $W_N^{(i)}$. Then, the polarized channel $W_{2^n}^{(i)}$ can be presented as $W_{p_1p_2...p_n}$.
	
	According to \eqref{eqn-5} and \eqref{eqn-6}, for binary-input memoryless symmetric AWGN channels, when GA method is used for polarization coding, the recursive process for the polarized channels satisfies the following
	\begin{equation}\label{eqn-17}
		m_{i+1}=\phi^{-1}\left(1-\left(1-\phi(m_i)\right)^2\right)\quad\text{if}\ p_{i+1}=0,
	\end{equation}
	\begin{equation}\label{eqn-18}
		m_{i+1}=2m_i\quad\text{if}\ p_{i+1}=1,
	\end{equation}
	where $\phi$ adopts the approximate function as shown in \eqref{eqn-7}.
	
	For the odd splitting channel, when $p_{i+1}=0$, it follows from \eqref{eqn-17} that
	\begin{equation}\label{eqn-19}
		\phi(m_{i+1})=2\phi(m_i)-\phi(m_i)^2\leq2\phi(m_i),
	\end{equation}
	\begin{equation}\label{eqn-20}
		\frac{\phi(m_{i+1})}{\phi(m_i)}\leq2.
	\end{equation}
	Since splitting channels with LLR mean greater than 10 are highly reliable and can be disregarded, when $m_i(\text{or}\ m_{i+1})\leq 10$, substituting $\phi(\gamma)\approx e^{a\gamma^c+b}$ from \eqref{eqn-7} yields
	\begin{align}
	\label{eqn-21} \frac{\mathrm{e}^{am_{i+1}^c+b}}{\mathrm{e}^{am_i^c+b}}&\leq2,\\
	\label{eqn-22} \mathrm{e}^{a(m_{i+1}^{c}-m_i^{c})}&\leq2,\\
	\label{eqn-23} {m_{i+1}}^{c}-{m_{i}}^{c}&\geq{\frac{1}{a}}\mathrm{ln}2,
	\end{align}
	Define the function $y=x^{c}$, then $x=y^{\frac1c}$. Define $\Delta y=y_{i+1}-y_{i}\geq\frac{1}{a}\mathrm{ln}2,\Delta x=x_{i+1}-x_{i}$. To analyze the properties of the function $x=y^{\frac1c}$, its first and second derivatives are calculated as
	\begin{equation}\label{eqn-24}
		\frac{dx}{dy}=\frac1cy^{\frac1c-1},
	\end{equation}
	\begin{equation}\label{eqn-25}
		\frac{d^2x}{dy^2}=\frac1c\left(\frac1c-1\right)y^{\frac1c-2}.
	\end{equation}
	The second derivative is greater than 0, indicating that the function $x=y^{\frac1c}$ is concave. This means that its tangent lines always lie below the function, and the slope of the tangent line is $k=\frac1cy^{\frac1c-1}$. Since polarization proceeds in the direction of decreasing LLR mean when ${p}_{i+1}=0$, we have
	\begin{equation}\label{eqn-26}
		\Delta x\geq\Delta x^{\prime}=k\Delta y\geq k\frac1a\mathrm{ln}2.
	\end{equation}
	When $y<10$, the following always holds
	\begin{equation}\label{eqn-27}
		k=\frac1cy^{\frac1c-1}\leq\frac1c10^{\frac1c-1},
	\end{equation}
	thus 
	\begin{equation}\label{eqn-28}
		\Delta x\geq\frac1c10^{\frac1c-1}\frac1a\ln2\approx-3.73\ln2.
	\end{equation}
	That is
	\begin{equation}\label{eqn-29}
		m_{i+1}-m_i\geq-3.73\ln2.
	\end{equation}
	
	Figure \ref{fig6} is a diagram of the tangent line to the function $x=y^{\frac1c}$ plotted for the case $y=8$, illustrating the calculation process described above.
	\begin{figure}[!t]
		\centering
		\includegraphics[width=3.2in]{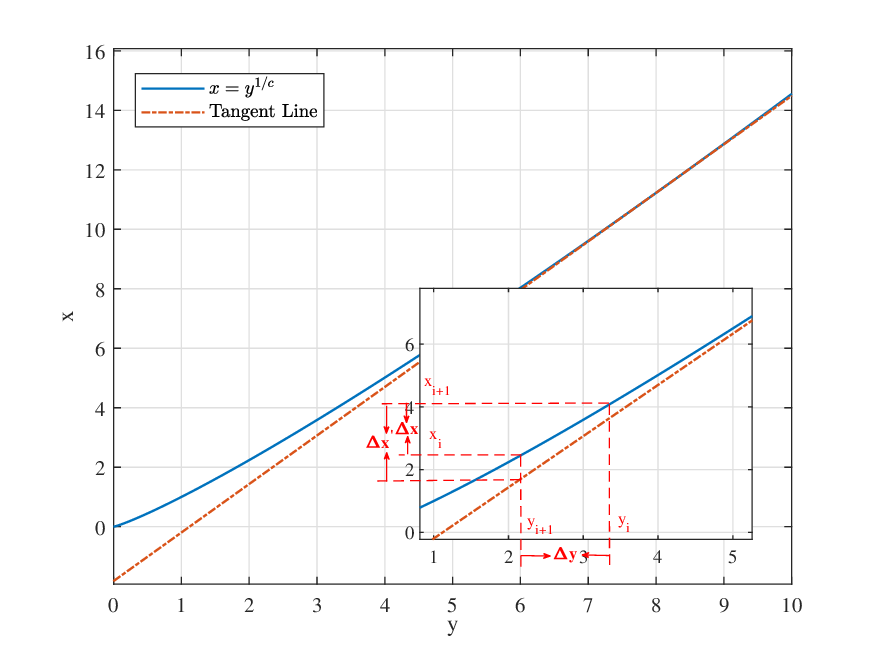}
		\caption{Diagram of the tangent line to function $x=y^{\frac1c}$ (illustrated for the case $y=8$)}
		\label{fig6}
	\end{figure}
	
	For the even splitting channel, when $p_{i+1}=1$, it follows from \eqref{eqn-18} that
	\begin{equation}\label{eqn-30}
		m_{i+1}=2m_i,
	\end{equation}
	\begin{equation}\label{eqn-31}
		m_{i+1}-m_i=m_i.
	\end{equation}
	For $\zeta\geq0$ and $s\geq0$, define
	\begin{equation}\label{eqn-32}
		T_s(\zeta)\triangleq\{\bar{p}\in\Omega:m_i\geq-3.73\ln\zeta,\forall i\geq s\},
	\end{equation}
	where $\bar{p}$	is the binary vectors belonging to the cylinder set $\Omega$. $T_s(\zeta)$ represents the set of splitting channels with LLR means greater than or equal to $-3.73\ \mathrm{ln}\zeta$ among all splitting channels whose subscript length is greater than or equal to	$s$. For $\bar{p}\in\mathcal{T}_s(\zeta)$ and $i\geq s$, we have
	\begin{equation}\label{eqn-33}
		m_{i+1}-m_i\geq-3.73\ln\zeta.
	\end{equation}
	
	From \eqref{eqn-29} and \eqref{eqn-33}, the following inequality is obtained.
	\begin{equation}\label{eqn-34}
		m_{i+1}-m_i \geq \begin{cases}-3.73\ln 2, & p_{i+1}=0 \\ -3.73\ln \zeta, & p_{i+1}=1\end{cases}
	\end{equation}
	Through recursion, the LLR mean of the splitting channels after the $n$-th polarization satisfies
	\begin{equation}\label{eqn-35}
		\begin{aligned}
			m_n \geq & -3.73 \ln 2-3.73 b_n \ln (\frac{\zeta}{2})+m_{n-1} \\
			\geq & -3.73 \ln 2-3.73 b_n \ln (\frac{\zeta}{2})-3.73 \ln 2 \\
			& -3.73 b_{n-1} \ln (\frac{\zeta}{2})+m_{n-2} \\
			\geq & -3.73 \ln 2-3.73 b_n \ln (\frac{\zeta}{2})-3.73 \ln 2-3.73 b_{n-1} \ln (\frac{\zeta}{2}) \\
			& -\cdots-3.73 \ln 2-3.73 b_{s+1} \ln (\frac{\zeta}{2})+m_s
		\end{aligned}
	\end{equation}
	Since $m_{s}\geq-3.73\ln\zeta$, \eqref{eqn-35} can be simplified to
	\begin{equation}\label{eqn-36}
		m_{n}\geq-3.73(n-s)\mathrm{ln}2-3.73\ \mathrm{ln}(\frac{\zeta}{2})\sum_{i=s+1}^{n}p_{i}-3.73\ln\zeta .
	\end{equation}
	For $n>s\geq0$ and $0<\eta<\frac{1}{2}$, define
	\begin{equation}\label{eqn-37}
		\mathcal{U}_{s,n}(\eta)\triangleq\left\{\bar{p}\in\Omega{:}\sum_{i=s+1}^np_i>(\frac{1}{2}-\eta)(n-s)\right\}.
	\end{equation}
	$\mathcal{U}_{s,n}(\eta)$ represents the set of splitting channels, among those with subscript lengths ranging from $s$ to $n$, where the number of trailing 1s in the subscript exceeds $\left(\frac12-\eta\right)(n-s)$. By substituting \eqref{eqn-37} into \eqref{eqn-36}, we obtain
	\begin{equation}\label{eqn-38}
		m_n\geq-3.73(n-s)\mathrm{ln}2-3.73\mathrm{ln}(\frac\zeta2)(\frac12-\eta)(n-s)-3.73\mathrm{ln}\zeta.
	\end{equation}
	Taking the values $\zeta=2^{-4}$ and $\eta=\frac1{20}$. Therefore, after the $n$-th polarization, when $\bar{p}\in\mathcal{T}_s(\zeta)\cap\mathcal{U}_{s,n}(\eta)$, the LLR mean of the splitting channels satisfies
	\begin{equation}\label{eqn-40}
		m_n\geq4.66n\ln2-4.66s\ln2+14.92\ln2.
	\end{equation}
	
	Next, we consider the transmission error probability. let $P\left(W_N^{(i)}\right)=P(\mathcal{C}_i)$ denote the BER of the $i$-th bit under the assumption that all previous bits have been decoded correctly. For a binary-input memoryless symmetric AWGN channel, $P(\mathcal{C}_n)$ can be calculated from the LLR mean as follows.
	\begin{equation}\label{eqn-41}
		P(C_n)=Q\left(\sqrt{\frac{m_n}2}\right),
	\end{equation}
	where $Q(x)=\frac1{\sqrt{2\pi}}\int_x^\infty e^{-\frac{t^2}2}dt$. Consider the upper bound of the Gaussian Q-function $Q(x)$ as follows\cite{tse2005fundamentals}.
	\begin{equation}\label{eqn-42}
		Q(x)<e^{-\frac{x^2}2},\quad x>1
	\end{equation}
	Then \eqref{eqn-41} can be further expressed as
	\begin{equation}\label{eqn-43}
		\begin{aligned}
			P(\mathcal{C}_n)&<e^{-\frac14(4.66n\ln2-4.66s\ln2+14.92\ln2)}\\&=2^{-\frac{2.33(n-s)}2-3.73},
		\end{aligned}
	\end{equation}
	indicating that when $\bar{p}\in\mathcal{T}_s(\zeta)\cap\mathcal{U}_{s,n}(\eta)$, BER of the splitting channels satisfies $P(\mathcal{C}_n)<2^{-\frac{2.33(n-s)}2-3.73}$.
	
	Next, we prove that \eqref{eqn-43} holds with high probability. From \cite[Lemma 1]{arikan2009channel}, the occurrence probability of set $\mathcal{T}_{s_{0}}(\zeta)$ is 
	\begin{equation}\label{eqn-44}
		P[T_{s_0}(\zeta)]\geq I_0-\frac\delta2,
	\end{equation}
	where $s_0$ is a finite integer, $\delta>0$, and $I_0$ is the channel capacity of the original binary-input memoryless symmetric AWGN channel. Using Chernoff's bound\cite{gallager1968information}, the occurrence probability of set $U_{s,n}(\eta)$ is
	\begin{equation}\label{eqn-45}
		P[U_{s,n}(\eta)]\geq1-2^{-(n-s)[1-H(1/2-\eta)]},
	\end{equation}
	where $H$ is the binary entropy function. A minimal $n_0$ can be found such that the right-hand side of \eqref{eqn-45} satisfies
	\begin{equation}\label{eqn-46}
		P[U_{s,n_0}(\eta)]\geq1-\frac\delta2,
	\end{equation}
	where $n_0$ is finite for any $s\geq0,0<\eta<1/2$ and $\delta>0$. Therefore, the occurrence probability of set $\mathcal{T}_{s_1}(\zeta_0)\cap\mathcal{U}_{s_1,n}(\eta_0)$ can be calculates as
	\begin{equation}\label{eqn-47}
		\begin{aligned}
			P[\mathcal{T}_{s_1}(\zeta_0)\cap\mathcal{U}_{s_1,n}(\eta_0)]&\geq(I_0-\frac\delta2)(1-\frac\delta2)\\
			&=I_0-\frac\delta2-\frac\delta2(I_0-\frac\delta2).
		\end{aligned}
	\end{equation}
	Since $I_0<1,I_0-\frac\delta2<1$, we have $-\frac\delta2(I_0-\frac\delta2)\geq\frac\delta2$, then
	\begin{equation}\label{eqn-48}
		P[\mathcal{J}_{s_1}(\zeta_0)\cap\mathcal{U}_{s_1,n}(\eta_0)]\geq I_0-\delta,\ n\geq n_1
	\end{equation}
	where $s_1$ is the minimal $s$ for given $\zeta_{0}$ and $\delta$, and $n_1$ is the minimal $n$ for given $s_1$,$\ \zeta_{0}$ and $\delta$.
	\vspace{2pt}
	
	Define $c\triangleq2^{\frac{2.33s_1}2-3.73}$ and
	\begin{equation}\label{eqn-49}
		\mathcal{V}_n\triangleq\left\{\bar{b}\in\Omega:P(\mathcal{C}_n)<c2^{-\frac{2.33n}2}\right\},\ n\geq0
	\end{equation}
	the sets have the relationship
	\begin{equation}\label{eqn-50}
		T_{s_1}(\zeta_0)\cap\mathcal{U}_{s_1,n}(\eta_0)\subset\mathcal{V}_n.\ n\geq n_1
	\end{equation}
	Therefore, for the occurrence probability, we have
	\begin{equation}\label{eqn-51}
		P[T_{s_1}(\zeta_0)\cap U_{s_1,n}(\eta_0)]<P(V_\mathrm{n}),
	\end{equation}
	then
	\begin{equation}\label{eqn-52}
		P(\mathcal{V}_{\mathrm{n}})\geq I_0-\delta.
	\end{equation}
	In addition, for $\mathcal{V}_{\mathrm{n}}$, we have
	\begin{equation}\label{eqn-53}
		P(\mathcal{V}_n)\:=\:\sum_{b_1^n\in X^n}\frac1{2^n}\mathbb{I}\Big\{P(W_{b_1^n})<c2^{-\frac{2.33n}2}\Big\}=\frac{|\mathcal{A}_N|}N,
	\end{equation}
	where $\mathbb{I}$ is the indicating function. $|\mathcal{A}_N|$ denotes the number of splitting channels satisfying the BER less than $c2^{-\frac{2.33n}2}$ and is defined as
	\begin{equation}\label{eqn-54}
		\mathcal{A}_N\triangleq\left\{i\in\{1\text{,...,}N\}:P(W_N^{(i)})<c2^{-\frac{2.33n}2}\right\}.
	\end{equation}
	Hence,
	\begin{equation}\label{eqn-55}
		\frac{|\mathcal{A}_{N}|}{N}\geq I_0-\delta,
	\end{equation}
	\begin{equation}\label{eqn-56}
		|\mathcal{A}_{N}|\geq N(I_0-\delta),
	\end{equation}
	where $I_0-\delta $ is a rate $R$ which is less than the channel capacity.
	
	Since $N=2^n$, $n=\log_2N$. Thus, we have proven that when GA is adopted for polar coding over a binary-input memoryless symmetric AWGN channel, for a given $N$, there always exists a set of splitting channels $|\mathcal{A}_N|\geq NR$ that are highly reliable, with reliability achieving $P(W_N^{(i)})<O(N^{-\frac{2.33}2})$. According to the isomorphism property of spatiotemporal polar codes in Lemma \ref*{le1}, it can be further proven that when spatiotemporal 2-D polar coding is implemented on multiple parallel AWGN channels, for given $T$ and $S$ satisfying $N=T\times S$, there always exists a set of splitting channels $|\mathcal{A}_{TS}|\geq TSR$ that are highly reliable, with reliability reaching $P(W_{TS}^{(i,j)})<O(N^{-\frac{2.33}2})$.

\end{IEEEproof}

\section{Simulation results and discussion}\label{sec5}
In this section, comparative simulations with the conventional 1-D time-domain polar coding are conducted to highlight the effectiveness and advantages of the proposed spatiotemporal 2-D polar coding scheme regarding reliability and latency. We consider a massive MIMO system, where the channel model is given in Section \ref{sec3} Part A. the code rate is set as 1/2, the polar decoder is the successive cancellation (SC) decoder, and the MIMO detector is the MMSE detector.

\begin{figure}[!t]
	\centering
	\includegraphics[width=3.2in]{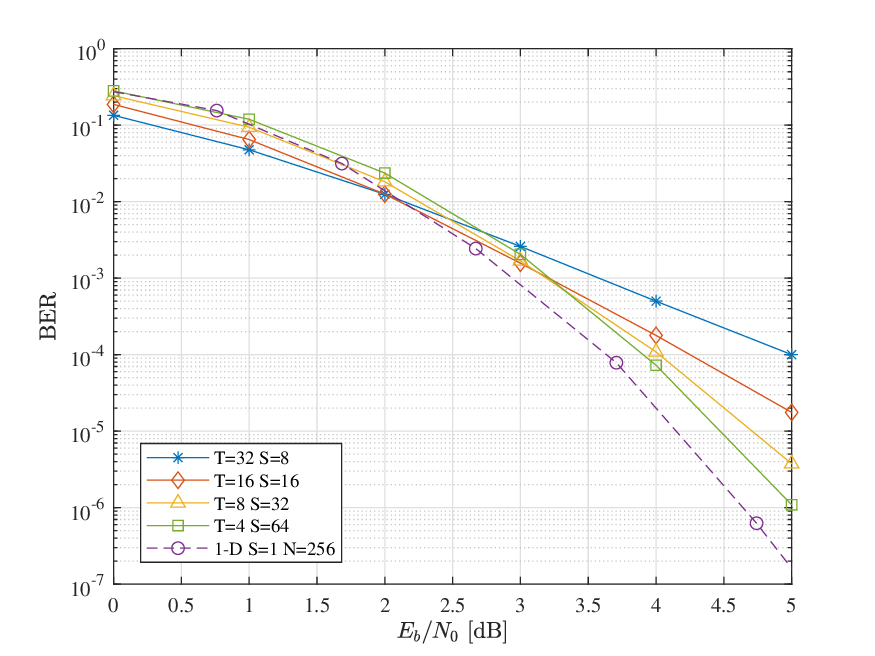}
	\caption{The BER of spatiotemporal 2-D polar codes under different space-time resource configurations ($\gamma=0.5$)}
	\label{fig7}
\end{figure}

Figure \ref{fig7} illustrates the BER performance comparison under different space-time resource configurations with the same 2-D total code length ($S\times T=256$). Simulation results demonstrate that the BER performance is similar across different configurations, a phenomenon that validates the space-time interchangeability characteristic of spatiotemporal polar codes, consistent with the theoretical analysis in Section \ref{sec3}. Notably, as the spatial dimension increases and the temporal dimension correspondingly decreases, the system performance shows a slight improvement. This enhancement can be attributed to the increased number of antennas, which leads to a more balanced SINR among channels under MMSE detection, thereby aligning more closely with the uniform channel assumption in this paper. Moreover, when the spatial dimension reaches a certain level, the system performance stabilizes. To further support this observation, we also include the BER performance of the conventional 1-D time-domain polar coding, using the same equivalent total code length $N=256$ and the same equivalent SINR as in the proposed 2-D scheme. It can be seen that this benchmark closely matches the stabilized performance of the 2-D scheme at high spatial dimensions, further confirming the validity and practicality of the uniform channel assumption in this study.

\begin{figure}[!t]
	\centering
	\includegraphics[width=3.2in]{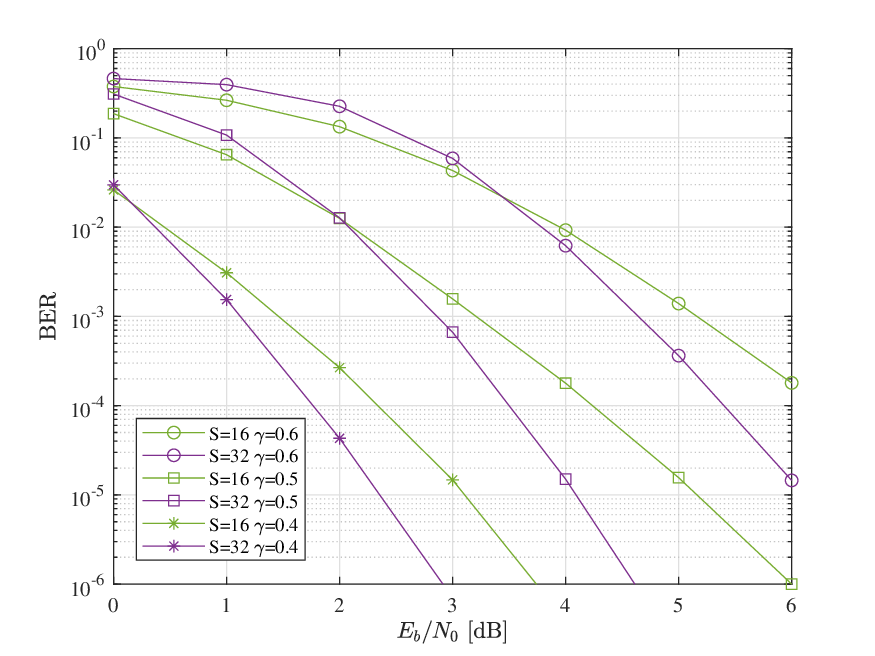}
	\caption{The BER of spatiotemporal 2-D polar codes under different $\gamma$ configurations ($T=16$)}
	\label{fig8}
\end{figure}

Figure \ref{fig8} shows the BER of the spatiotemporal polar codes under different $\gamma=\frac{S}{L}$ configurations with $S=16$ and $32$. It can be seen that the BER performance of spatiotemporal polar coding all deteriorates as $\gamma$ increases (i.e. $L$ decreases). According to \eqref{normal_dist}, the increase in $\gamma$ causes the mean value of $\mathrm{SINR}_k$ to decrease, which can be numerically verified since $0 \le \mu_{\gamma}, \sigma_{\gamma}^2 \le 1$. The decrease in $\mathbb{E}(\mathrm{SINR}_k)$ reduces the SINR and degrades the BER performance. On the other hand, a reduction in the number of receiving antennas means a reduction in spatial DoF, resulting in increased interference and thus deteriorated decoding performance. This reveals that, to meet the requirement of URLLC, the reliability can be guaranteed by increasing $L$ when $S$ is fixed, and the effectiveness of the proposed spatiotemporal 2-D polar coding scheme for massive MIMO systems with MMSE receivers is also verified.

\begin{figure*}[!t]
	\centering
	\subfigure[$N=T=16$]{
		\label{fig9a}
		\includegraphics[width=0.32\textwidth]{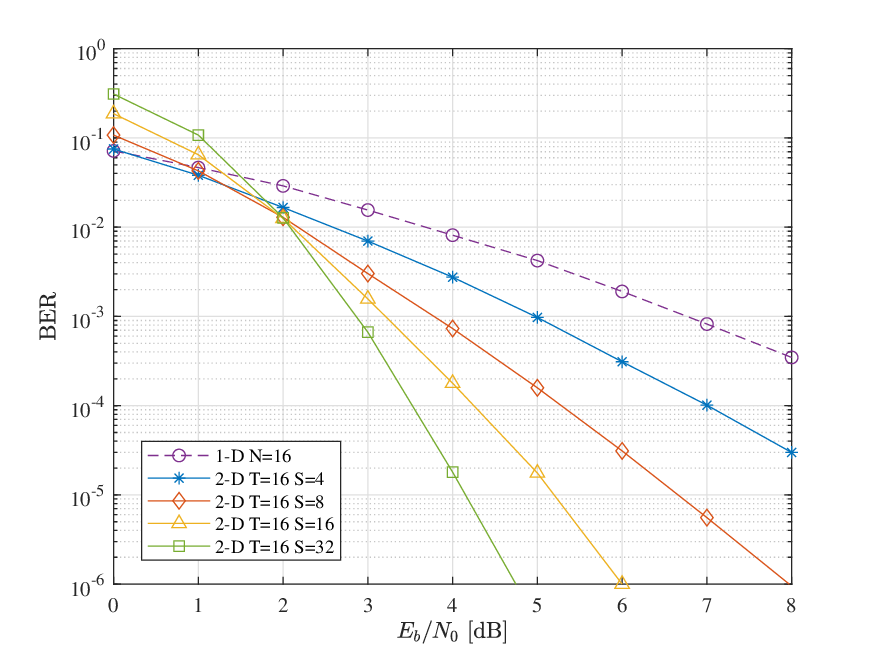}}
	\subfigure[$N=T=32$]{
		\label{fig9b}
		\includegraphics[width=0.32\textwidth]{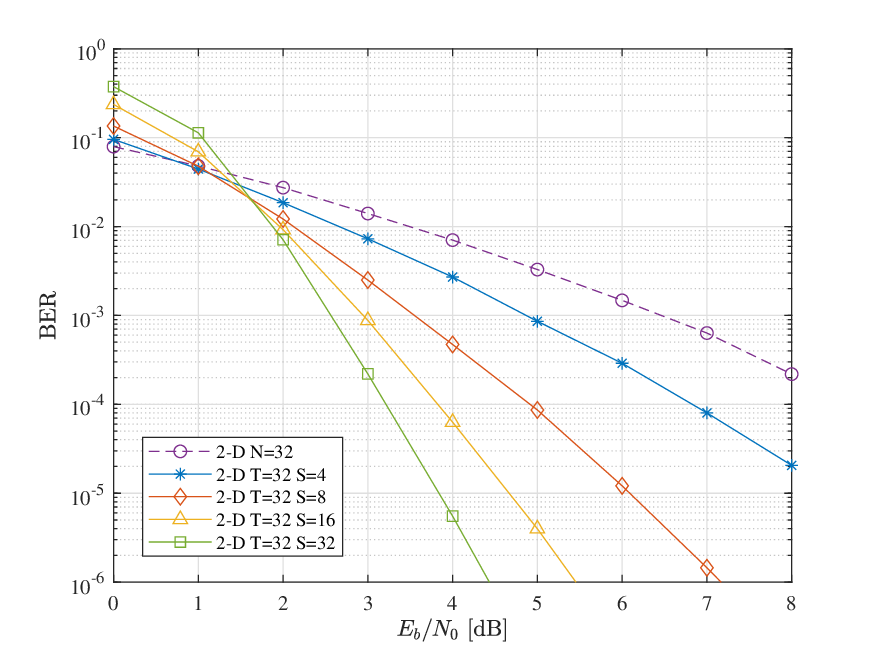}}
	\subfigure[$N=T=64$]{
		\label{fig9c}
		\includegraphics[width=0.32\textwidth]{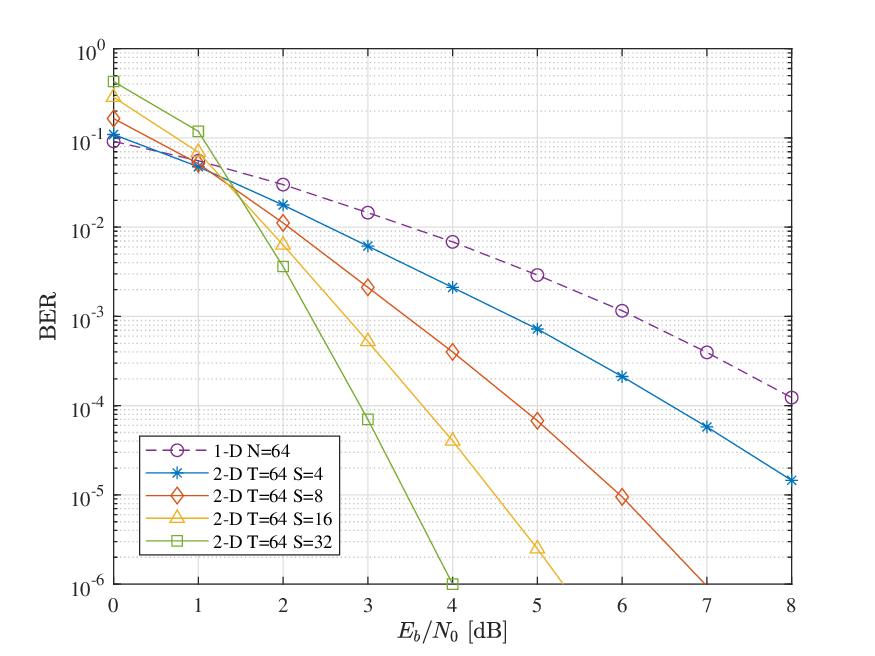}}
	\caption{The BLER of spatiotemporal 2-D polar codes and 1-D polar codes under identical time-domain blocklengths ($\gamma=0.5$)}
	\label{fig9}
\end{figure*}
\begin{figure*}[!t]
	\centering
	\subfigure[$N=S \times T=32$]{
		\label{fig10a}
		\includegraphics[width=0.32\textwidth]{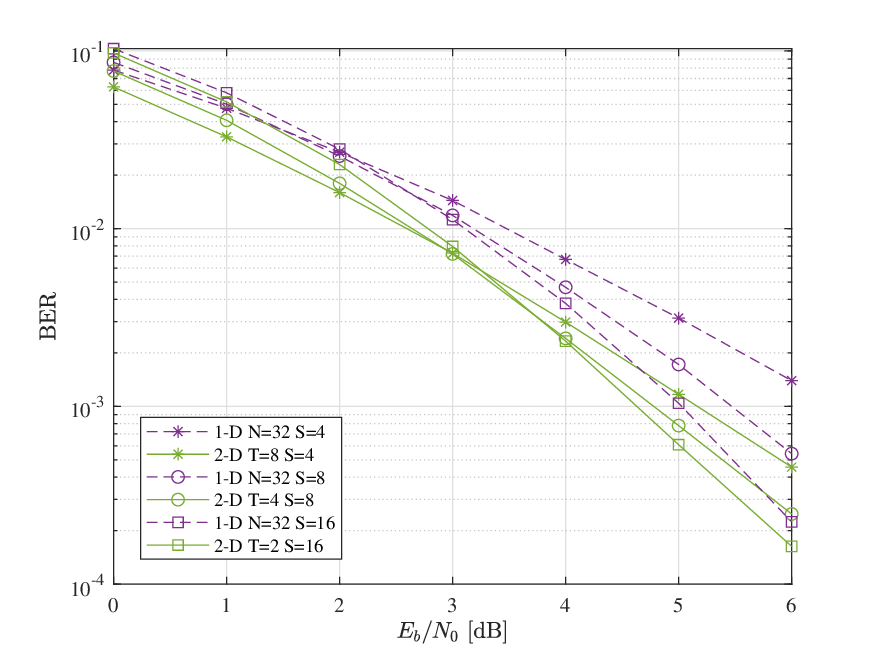}}
	\subfigure[$N=S \times T=64$]{
		\label{fig10b}
		\includegraphics[width=0.32\textwidth]{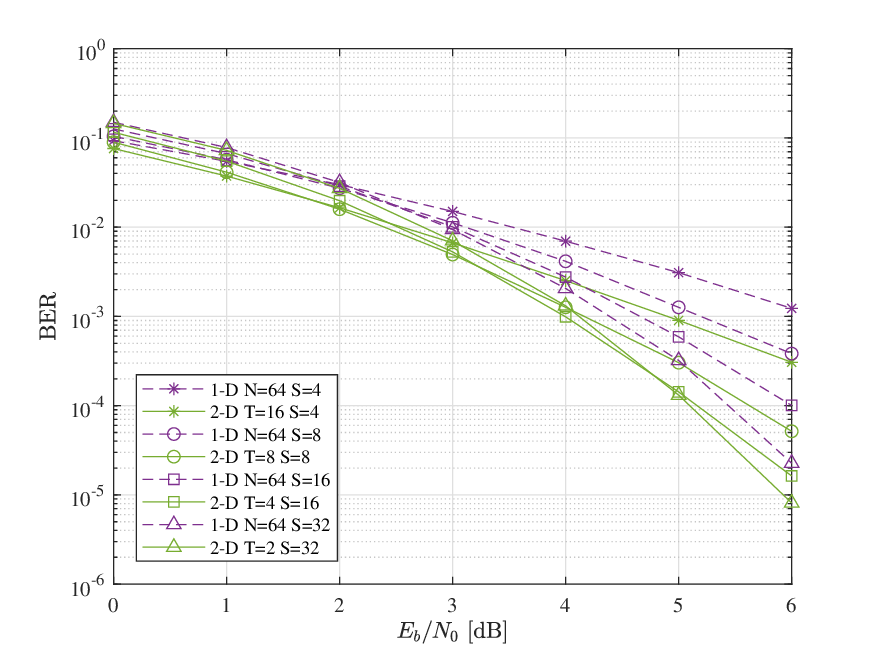}}
	\subfigure[$N=S \times T=128$]{
		\label{fig10c}
		\includegraphics[width=0.32\textwidth]{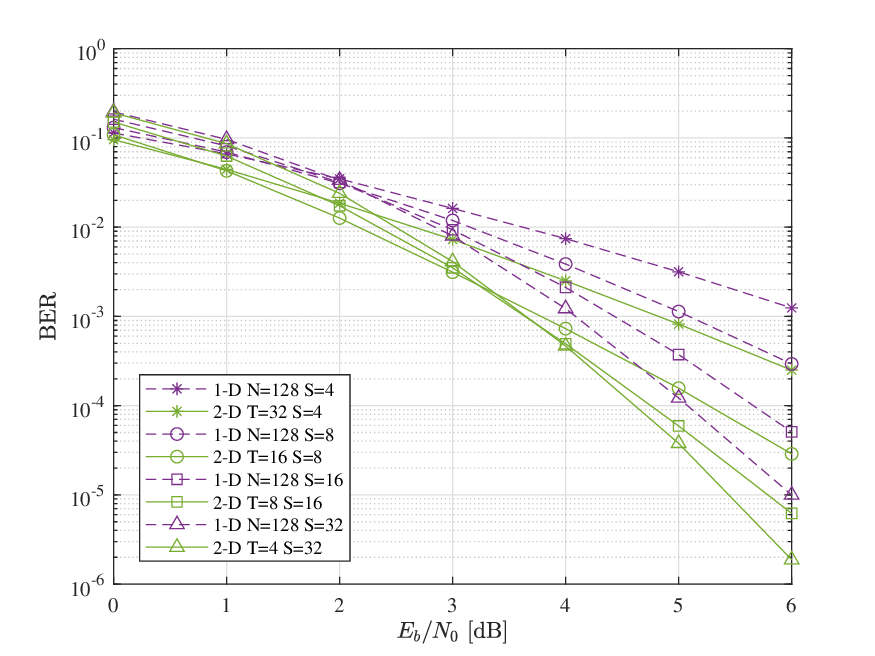}}
	\caption{The FER of spatiotemporal 2-D polar codes and 1-D polar codes under identical equivalent total code length ($\gamma=0.5$)}
	\label{fig10}
\end{figure*}

Figure \ref{fig9} presents the performance comparison between the proposed spatiotemporal 2-D polar codes and conventional 1-D time-domain polar codes under identical time-domain blocklengths of $N=T=16, 32$, and $64$. As observed, conventional 1-D polar coding exhibits limited performance in MIMO systems, particularly under short time-domain blocklengths, due to its inability to exploit spatial DoF. In contrast, the proposed spatiotemporal 2-D polar codes exhibit consistently better BER performance across all blocklengths by utilizing spatial coding among multiple bit-streams. Furthermore, the performance gain becomes more pronounced with increasing spatial dimension. For instance, for a short length of $N=32$, the performance gain of 2-D polar coding with $S=8$ is more than 3dB at BER = $10^{-4}$. These results clearly demonstrate that the proposed 2-D polar coding scheme achieves significant BER improvements over traditional time-domain polar coding in massive MIMO systems, while maintaining the same time-domain latency---particularly under short blocklength scenarios.

Figure \ref{fig10} presents the performance comparison between the proposed spatiotemporal 2-D polar codes and conventional 1-D time-domain polar codes under identical equivalent total code length, with values set to $N=S\times T=32, 64$, and $128$, respectively. According to the theoretical analysis in Section \ref*{sec3}, both schemes are expected to achieve similar BER performance under the same equivalent total code length. As shown in the figure, the simulation results confirm this expectation, and the proposed 2-D scheme further outperforms the 1-D counterpart, primarily because the spatial encoding across bit-streams in the 2-D scheme enables a more flexible distribution of bits over varying channel conditions, effectively mitigating the impact of poor channels that would otherwise dominate entire blocks in the 1-D case. Moreover, as the spatial dimension increases, the performance of the 2-D scheme tends to stabilize, which is consistent with the results observed in Figure \ref{fig7}. 

Next, we present a latency analysis comparing the two schemes. Although the current system model is based on an ideal baseband transmission link, to better illustrate the latency advantage of the proposed spatiotemporal 2-D polar codes in practical communication systems, we refer to the frame structure parameters defined in the 5G New Radio (NR) standard and perform an analogical estimation of transmission delay. In 5G NR, the minimum scheduling unit at the physical layer is a slot, which typically contains 14 orthogonal frequency division multiplexing (OFDM) symbols. Taking a 15 kHz subcarrier spacing as an example, one slot corresponds to a transmission duration of 1 ms. For simplicity, we assume that one BPSK-modulated symbol occupies the same transmission duration as one OFDM symbol in the physical layer, i.e., 1/14 ms, enabling an intuitive mapping of symbol-level latency to real-world time slots. Under such a comparison framework, suppose the equivalent total code length of the transmitted polar codes is $N$. The transmission delays of the two schemes can then be estimated as follows.
\begin{itemize}
	\item \textbf{Traditional 1-D time-domain polar coding schemes:}
	
	Due to the serial transmission and decoding dependency in 1-D polar codes, decoding must wait until all $N$ symbols have been received. Assuming the symbols are sent sequentially in the time domain, the required number of time slots is approximately
	\begin{equation}
		T_{1\mathrm{D}}=\lceil\frac N{14}\rceil. 
	\end{equation}
	The transmission delay is estimated as
	\begin{equation}
		\mathrm{Latency}_{1\mathrm{D}}=T_{1\mathrm{D}}\times1\mathrm{~ms}.
	\end{equation}
	\item \textbf{Spatiotemporal 2-D polar coding schemes:}
	
	The proposed 2-D coding structure allows parallel encoding and transmission over $S$ spatial streams, while performing time-domain coding with blocklength $T$, satisfying $N = S \times T$. Due to the spatial parallelism, the delay is primarily determined by the time-domain length $T$. Therefore, the number of required time slots is
	\begin{equation}
		T_{2\mathrm{D}}=\lceil\frac T{14}\rceil=\lceil\frac{N/S}{14}\rceil.
	\end{equation}
	The estimated transmission delay is
	\begin{equation}
		\mathrm{Latency}_{{2\mathrm{D}}}=T_{{2\mathrm{D}}}\times1\mathrm{~ms}.
	\end{equation}
\end{itemize}

Thus, the proposed 2-D scheme reduces transmission delay by approximately a factor of $S$ compared to 1-D scheme, while maintaining comparable reliability. For example, as shown in Figure \ref{fig10c}, when the total code length is $N=S\times T=128$ with $S=4$, the proposed scheme achieves a performance gain of around 1dB at BER=$10^{-3}$. More importantly, the required number of time slots in the 1-D scheme is $\lceil128/14\rceil=10$, resulting in a latency of approximately 10 ms. In contrast, the 2-D scheme only requires $\lceil(128/4)/14\rceil=\lceil32/14\rceil=3$ time slots, leading to a latency of around 3 ms, effectively reducing transmission delay by over a factor of 3. It should be noted that in practical communication systems, end-to-end latency is also affected by factors such as channel scheduling, encoding and decoding time, and feedback mechanisms. However, at the physical layer, the proposed 2-D polar coding scheme significantly reduces waiting time by leveraging spatial DoF, thereby enhancing low-latency performance. These results demonstrate that, under the same equivalent code length, the proposed 2-D polar coding scheme achieves comparable or even superior BER performance compared to conventional 1-D polar coding, while reducing latency by a factor of $S$. This highlights the potential of the proposed scheme to reduce latency without sacrificing reliability, making it particularly attractive for low-latency communications in massive MIMO systems.


\section{Conclusion}\label{sec6}
This paper proposes an innovative spatiotemporal 2-D polar coding scheme for massive MIMO systems, which effectively leverages spatial degrees of freedom and the inherent advantages of polar codes to reduce latency. A unified theoretical coding framework is established, enabling separate polar transforms in both the spatial and temporal domains. Furthermore, the structural isomorphism among different 2-D structures is formally demonstrated. Based on the constructive Gaussian approximation, the proposed scheme is theoretically proven to approach channel capacity under finite blocklength constraints. Simulation results further validate its superiority over conventional 1-D time-domain polar codes, demonstrating that it can reduce latency under the same reliability requirements, or alternatively, enhance reliability under the same latency constraint---highlighting its strong potential for both theoretical research and practical deployment in future 6G eURLLC transmission.


\ifCLASSOPTIONcaptionsoff
  \newpage
\fi

\bibliographystyle{IEEEtran}
\bibliography{main}

\begin{thebibliography}{10}
\providecommand{\url}[1]{#1}
\csname url@samestyle\endcsname
\providecommand{\newblock}{\relax}
\providecommand{\bibinfo}[2]{#2}
\providecommand{\BIBentrySTDinterwordspacing}{\spaceskip=0pt\relax}
\providecommand{\BIBentryALTinterwordstretchfactor}{4}
\providecommand{\BIBentryALTinterwordspacing}{\spaceskip=\fontdimen2\font plus
\BIBentryALTinterwordstretchfactor\fontdimen3\font minus
  \fontdimen4\font\relax}
\providecommand{\BIBforeignlanguage}[2]{{%
\expandafter\ifx\csname l@#1\endcsname\relax
\typeout{** WARNING: IEEEtran.bst: No hyphenation pattern has been}%
\typeout{** loaded for the language `#1'. Using the pattern for}%
\typeout{** the default language instead.}%
\else
\language=\csname l@#1\endcsname
\fi
#2}}
\providecommand{\BIBdecl}{\relax}
\BIBdecl

\bibitem{schulz2017latency}
P.~Schulz, M.~Matthe, H.~Klessig, M.~Simsek, G.~Fettweis, J.~Ansari, S.~A.
  Ashraf, B.~Almeroth, J.~Voigt, I.~Riedel \emph{et~al.}, ``Latency critical
  iot applications in {5G}: Perspective on the design of radio interface and
  network architecture,'' \emph{IEEE Communications Magazine}, vol.~55, no.~2,
  pp. 70--78, 2017.

\bibitem{sutton2019enabling}
G.~J. Sutton, J.~Zeng, R.~P. Liu, W.~Ni, D.~N. Nguyen, B.~A. Jayawickrama,
  X.~Huang, M.~Abolhasan, Z.~Zhang, E.~Dutkiewicz \emph{et~al.}, ``Enabling
  technologies for ultra-reliable and low latency communications: From {PHY}
  and {MAC} layer perspectives,'' \emph{IEEE Communications Surveys \&
  Tutorials}, vol.~21, no.~3, pp. 2488--2524, 2019.

\bibitem{sachs20185g}
J.~Sachs, G.~Wikstrom, T.~Dudda, R.~Baldemair, and K.~Kittichokechai, ``{5G}
  radio network design for ultra-reliable low-latency communication,''
  \emph{IEEE network}, vol.~32, no.~2, pp. 24--31, 2018.

\bibitem{you2023toward}
X.~You, Y.~Huang, S.~Liu, D.~Wang, J.~Ma, C.~Zhang, H.~Zhan, C.~Zhang,
  J.~Zhang, Z.~Liu \emph{et~al.}, ``Toward {6G} {TK}$\mu$ extreme connectivity:
  Architecture, key technologies and experiments,'' \emph{IEEE Wireless
  Communications}, vol.~30, no.~3, pp. 86--95, 2023.

\bibitem{saad2019vision}
W.~Saad, M.~Bennis, and M.~Chen, ``A vision of {6G} wireless systems:
  Applications, trends, technologies, and open research problems,'' \emph{IEEE
  network}, vol.~34, no.~3, pp. 134--142, 2019.

\bibitem{1998Constrained}
D.~E. Lazic, T.~Beth, and S.~Egner, ``Constrained capacity of the {AWGN}
  channel,'' in \emph{Information Theory, 1998. Proceedings. 1998 IEEE
  International Symposium on}, 1998.

\bibitem{2007A}
J.~Shi and R.~D. Wesel, ``A study on universal codes with finite block
  lengths,'' \emph{IEEE Transactions on Information Theory}, vol.~53, no.~9,
  pp. 3066--3074, 2007.

\bibitem{2010Channel}
Y.~Polyanskiy, H.~V. Poor, and S.~Verdu, ``Channel coding rate in the finite
  blocklength regime,'' \emph{IEEE Transactions on Information Theory},
  vol.~56, no.~5, pp. 2307--2359, 2010.

\bibitem{you20236g}
X.~You, ``{6G} extreme connectivity via exploring spatiotemporal
  exchangeability,'' \emph{Science China Information Sciences}, vol.~66, no.~3,
  p. 130306, 2023.

\bibitem{you2023closed}
X.~You, B.~Sheng, Y.~Huang, W.~Xu, C.~Zhang, D.~Wang, P.~Zhu, and C.~Ji,
  ``Closed-form approximation for performance bound of finite blocklength
  massive {MIMO} transmission,'' \emph{IEEE Transactions on Communications},
  vol.~71, no.~12, pp. 6939--6951, 2023.

\bibitem{hoydis2011iterative}
J.~Hoydis, R.~Couillet, and M.~Debbah, ``Iterative deterministic equivalents
  for the performance analysis of communication systems,'' \emph{arXiv preprint
  arXiv:1112.4167}, 2011.

\bibitem{li2005distribution}
P.~Li, D.~Paul, R.~Narasimhan, and J.~Cioffi, ``On the distribution of {SINR}
  for the {MMSE} {MIMO} receiver and performance analysis,'' \emph{IEEE
  Transactions on Information Theory}, vol.~52, no.~1, pp. 271--286, 2005.

\bibitem{kammoun2009central}
A.~Kammoun, M.~Kharouf, W.~Hachem, and J.~Najim, ``A central limit theorem for
  the {SINR} at the {LMMSE} estimator output for large-dimensional signals,''
  \emph{IEEE Transactions on Information Theory}, vol.~55, no.~11, pp.
  5048--5063, 2009.

\bibitem{moustakas2013sinr}
A.~L. Moustakas and P.~Kazakopoulos, ``{SINR} statistics of correlated {MIMO}
  linear receivers,'' \emph{IEEE transactions on information theory}, vol.~59,
  no.~10, pp. 6490--6500, 2013.

\bibitem{you2022spatiotemporal}
X.~You, C.~Zhang, B.~Sheng, Y.~Huang, C.~Ji, Y.~Shen, W.~Zhou, and J.~Liu,
  ``Spatiotemporal {2-D} channel coding for very low latency reliable {MIMO}
  transmission,'' in \emph{2022 IEEE Globecom Workshops (GC Wkshps)}.\hskip 1em
  plus 0.5em minus 0.4em\relax IEEE, 2022, pp. 473--479.

\bibitem{arikan2009channel}
E.~Arikan, ``Channel polarization: A method for constructing capacity-achieving
  codes for symmetric binary-input memoryless channels,'' \emph{IEEE
  Transactions on information Theory}, vol.~55, no.~7, pp. 3051--3073, 2009.

\bibitem{korada2009polar}
S.~B. Korada, ``Polar codes for channel and source coding,'' Ph.D.
  dissertation, EPFL, 2009.

\bibitem{trifonov2012efficient}
P.~Trifonov, ``Efficient design and decoding of polar codes,'' \emph{IEEE
  transactions on communications}, vol.~60, no.~11, pp. 3221--3227, 2012.

\bibitem{wu2014construction}
D.~Wu, Y.~Li, and Y.~Sun, ``Construction and block error rate analysis of polar
  codes over {AWGN} channel based on {G}aussian approximation,'' \emph{IEEE
  Communications Letters}, vol.~18, no.~7, pp. 1099--1102, 2014.

\bibitem{you2021towards}
X.~You, C.-X. Wang, J.~Huang, X.~Gao, Z.~Zhang, M.~Wang, Y.~Huang, C.~Zhang,
  Y.~Jiang, J.~Wang \emph{et~al.}, ``Towards {6G} wireless communication
  networks: Vision, enabling technologies, and new paradigm shifts,''
  \emph{Science China information sciences}, vol.~64, pp. 1--74, 2021.

\bibitem{rasti2022evolution}
M.~Rasti, S.~K. Taskou, H.~Tabassum, and E.~Hossain, ``Evolution toward {6G}
  multi-band wireless networks: A resource management perspective,'' \emph{IEEE
  Wireless Communications}, vol.~29, no.~4, pp. 118--125, 2022.

\bibitem{chung2001analysis}
S.-Y. Chung, T.~J. Richardson, and R.~L. Urbanke, ``Analysis of sum-product
  decoding of low-density parity-check codes using a {G}aussian
  approximation,'' \emph{IEEE Transactions on Information theory}, vol.~47,
  no.~2, pp. 657--670, 2001.

\bibitem{poor1997probability}
H.~V. Poor and S.~Verd{\'u}, ``Probability of error in {MMSE} multiuser
  detection,'' \emph{IEEE transactions on Information theory}, vol.~43, no.~3,
  pp. 858--871, 1997.

\bibitem{tse2000linear}
D.~N.~C. Tse and O.~Zeitouni, ``Linear multiuser receivers in random
  environments,'' \emph{IEEE Transactions on Information Theory}, vol.~46,
  no.~1, pp. 171--188, 2000.

\bibitem{verdu1998multiuser}
S.~Verd{\'u}, \emph{Multiuser detection}.\hskip 1em plus 0.5em minus
  0.4em\relax Cambridge university press, 1998.

\bibitem{paulraj2003introduction}
A.~Paulraj, R.~Nabar, and D.~Gore, \emph{Introduction to space-time wireless
  communications}.\hskip 1em plus 0.5em minus 0.4em\relax Cambridge university
  press, 2003.

\bibitem{doan2018decoding}
N.~Doan, S.~A. Hashemi, M.~Mondelli, and W.~J. Gross, ``On the decoding of
  polar codes on permuted factor graphs,'' in \emph{2018 IEEE Global
  Communications Conference (GLOBECOM)}.\hskip 1em plus 0.5em minus 0.4em\relax
  IEEE, 2018, pp. 1--6.

\bibitem{geiselhart2021automorphism}
M.~Geiselhart, A.~Elkelesh, M.~Ebada, S.~Cammerer, and S.~ten Brink, ``On the
  automorphism group of polar codes,'' in \emph{2021 IEEE International
  Symposium on Information Theory (ISIT)}.\hskip 1em plus 0.5em minus
  0.4em\relax IEEE, 2021, pp. 1230--1235.

\bibitem{tse2005fundamentals}
D.~Tse and P.~Viswanath, \emph{Fundamentals of wireless communication}.\hskip
  1em plus 0.5em minus 0.4em\relax Cambridge university press, 2005.

\bibitem{gallager1968information}
R.~G. Gallager, \emph{Information theory and reliable communication}.\hskip 1em
  plus 0.5em minus 0.4em\relax Springer, 1968, vol. 588.

\end{thebibliography}

\end{document}